\documentclass[letterpaper,twocolumn,10pt]{article}
\usepackage{usenix2021_SOUPS}
\usepackage{balance}
\usepackage{wasysym}
\usepackage{xcolor}         

\usepackage{enumitem}
\usepackage{mathptmx}
\usepackage{graphicx}
\usepackage{array}
\usepackage{multirow}
\usepackage{pifont}
\usepackage{color}
\usepackage{siunitx}
\usepackage{booktabs}
\usepackage{rotating}
\usepackage{xparse}
\usepackage{blindtext}
\usepackage{tabularx}
\usepackage{url}
\usepackage{cite}
\usepackage{amssymb}
\usepackage{pifont}
\usepackage[font=small]{caption}
\usepackage{subcaption}
\usepackage{tikz}
\usepackage{makecell}

\renewcommand{\paragraph}[1]{\vspace{5pt}\noindent\textbf{#1\quad}}

\usepackage{soul}
\usepackage{xspace}

\newcommand{\comment}[1]{\relax}
\newcommand{\eg}{e.g.,\@\xspace}
\newcommand{\ie}{i.e.,\@\xspace}

\newcolumntype{R}{>{\raggedleft\arraybackslash}X}
\newcolumntype{L}{>{\raggedright\arraybackslash}X}
\newcommand{\uniqueraters}{17,280\xspace}
\newcommand{\uniquecomments}{107,620\xspace}

\usepackage{textcomp}
\usepackage[utf8]{inputenc}
\DeclareUnicodeCharacter{2212}{\textminus}

\usepackage{csquotes}
\renewenvironment{displayquote}{%
   \list{}{%
     \fontsize{9.2pt}{10.7pt}\selectfont
     \leftmargin0.32cm   
     \rightmargin0.32cm
   }
   \item\relax
}
{\endlist}

\usepackage{adjustbox}

\newif\ifext\extfalse

\begin{document}
\pagestyle{plain}
\thispagestyle{empty}

\title{Designing Toxic Content Classification for a Diversity of Perspectives}
\newif\ifblind\blindfalse 

\def\uiuc{$^\dagger$}
\def\umich{$^\triangleleft$}
\def\gtech{$^\diamond$}
\def\Akamai{$^\ddagger$}
\def\cloudflare{$^\triangleright$}
\def\google{$^\circ$}
\def\merit{$^\S$}
\def\authspace{\hspace{10pt}}

\def\plainauthor{Author name(s) for PDF metadata.}

\def\stanford{$^\triangleleft$}
\def\uiuc{$^\dagger$}
\def\google{$^\circ$}
\def\authspace{\hspace{10pt}}

\author{
    {\rm Deepak Kumar\stanford \authspace
    Patrick Gage Kelley\google \authspace
    Sunny Consolvo\google \authspace
    Joshua Mason\uiuc \authspace
    Elie Bursztein\google}
    \and
    {\rm Zakir Durumeric\stanford \authspace
    Kurt Thomas\google \authspace
    Michael Bailey\uiuc}
    \and
    \stanford Stanford University\authspace
    \google Google \authspace
    \uiuc University of Illinois at Urbana-Champaign
}

\maketitle

\begin{abstract}
In this work, we demonstrate how existing classifiers for identifying toxic
comments online fail to generalize to the diverse concerns of Internet
users. We survey 17,280~participants to understand how user expectations for
what constitutes toxic content differ across demographics, beliefs, and
personal experiences.  We find that groups historically at-risk of
harassment---such as people who identify as LGBTQ+ or young adults---are
more likely to to flag a random comment drawn from Reddit, Twitter, or 4chan
as toxic, as are people who have personally experienced harassment in the
past. Based on our findings, we show how current one-size-fits-all toxicity
classification algorithms, like the Perspective API from Jigsaw, can improve
in accuracy by 86\% on average through personalized model tuning.
Ultimately, we highlight current pitfalls and new design directions that can
improve the equity and efficacy of toxic content classifiers for all users.
\end{abstract}

\section{Introduction}
Online hate and harassment is a pernicious threat facing 48\% of Internet
users~\cite{thomas2021hate}. In response to this growing challenge, online
platforms have developed automated tools to take action against toxic content
(e.g., hate speech, threats, identity attacks). Examples include Yahoo's abusive
language classifier trained on crowdsourced labels attached to news
comments~\cite{nobata2016abusive}, Google Jigsaw's Perspective API, which is
trained on Wikipedia moderation verdicts for abuse as well as samples from other
online communities~\cite{wulczyn2017ex, jigsaw-kaggle}, and Instagram's recent
classifier that detects harassing comments posted as a reply to
photos~\cite{instagram-bullying}.

Although platforms have used these classifiers to address toxic content in
direct violation of their policies~\cite{twitter-flags}, a variety of content
that is not toxic enough to violate policy may still cause harm to Internet
users~\cite{twitter-still-toxic}. These ``gray areas'' stem from the fact that
users may disagree about what constitutes toxic content online based on their
lived experiences, cultural perspective, political views towards free speech, or
access to appropriate context~\cite{sambasivan2019they, gualdo2015emotional}.
While prior research has demonstrated that certain groups are more at-risk of
experiencing online hate and harassment~\cite{pew2017harassment,
thomas2021hate}, no study has investigated how users from diverse backgrounds
interpret online toxicity or how their views on what content they would like to
see online differ. Understanding these nuanced differences is an important first
step to designing harassment defenses for diverse Internet users.


In this work, we investigate divergent user interpretations of toxic content and
identify whether current classifiers can be tuned to accommodate a diversity of
perspectives. At the core of our study, we develop a survey instrument that asks
\uniqueraters~participants to rate and label the toxicity of 20~random comments
drawn from \uniquecomments~Twitter, Reddit, and 4chan. In tandem, we collect
demographic data and log participant's previous exposure and experiences with
online harassment. Taken together, our survey instrument provides access to a
diverse set of perspectives on why people deem certain comments as toxic. We
explore this data in three steps: we investigate user ratings of toxic content
in aggregate, we identify the factors that result in identical comments
receiving divergent ratings, and finally, we demonstrate how modern classifiers
can be tuned to better accommodate differing user perspectives.



Participants frequently disagree on whether comments were toxic. In aggregate,
participants labeled 53\% of our dataset as ``not toxic'', 39\% as ``slightly''
or ``moderately toxic'' and the remaining 8\% as ``very'' or ``extremely
toxic''. However, 85\% of comments exhibited some form of disagreement,
including whether participants were comfortable seeing the comment on any online
platform. Even when participants uniformly agree that a comment was toxic, they
disagree about the subcategory the comment belonged to (\eg a threat versus an
insult). As such, a la carte models that isolate individual classes of toxic
content---for instance, identity-based attacks~\cite{warner2012detecting}---may
fail to adequately meet the needs of a user base with diverse perspectives on
toxic content.

%

A variety of factors influence how users perceive toxicity. We find that a
participant's personal experience with harassment, whether the participant
belongs to an at-risk group frequently targeted by
harassment~\cite{pew2017harassment,datasociety2016harassment}, and a
participant's attitudes towards filtering online discourse all correlate with
rating a comment as toxic or not. For example, holding all other factors
constant, the odds that a participant rates
a comment as toxic increase 1.64~times if they identify as LGBTQ+.
Alternatively, these odds decrease by 0.78~times for users who regularly witness
others targeted by toxic content, potentially due to desensitization. Combined,
no single demographic variable or experience defines how participants interpret
toxic content, underscoring the need for diverse raters in data labeling and
model construction.

\looseness=-1
Finally, we investigate how we might leverage current state-of-the-art
classifiers to enable diverse user perspectives of toxic content online. We
focus on Jigsaw's Perspective API~\cite{perspective-api} and Instagram's comment
nudge~\cite{instagram-bullying}. As a baseline, we find for content that
Perspective deemed 90\% likely to be toxic, only 50\% of our participants
agreed. Similarly, Instagram's classifier flagged only 27\% of comments that a majority of
our participants rated as toxic. We propose potential improvements based on
\emph{personalized tuning}---finding a threshold for the classifier that is set
based on individual responses or in larger demographic groups. These
improvements achieve a 86\% boost in accuracy per individual and a 22\%
improvement in accuracy per demographic cohort, highlighting personalized
modeling as a future direction in toxicity classification.

%
%

We conclude with a discussion of how to overcome the limitations of crowdsourced
labeling and one-size-fits-all classification that we identified through our
work. To this end, we have shared our results with Jigsaw and have released our
labeled
dataset\footnote{\url{https://data.esrg.stanford.edu/study/toxicity-perspectives}}
to enable other researchers to reproduce our analysis, build new classifiers,
and further explore how different individuals perceive toxic behavior online.

\section{Background \& Related Work}
\label{sec:background}

\subsection{What is toxic content?}
We use the term \emph{toxic content} as an umbrella for identity-based attacks
such as anti-Semitism or racism posted publicly to social
media~\cite{adl-antisemitism, warner2012detecting, zannettou2019characterizing},
bullying in online gaming or replies to posts~\cite{sambasivan2019they,
kwak2015exploring},
trolling~\cite{cheng2015antisocial}, threats of violence, sexual harassment, and
more~\cite{redmiles2019just, thomas2021hate}. These attacks represent just a subset of abuse
stemming from \emph{hate and harassment}, a much broader threat that encompasses
any activity where an attacker attempts to inflict emotional harm on a target
(\eg stalking, doxxing, sextortion, and intimate partner
violence)~\cite{thomas2021hate, citron2014addressing}. Unlike spam, phishing, or related abuse
classification problems that can rely on expert raters, toxic content is an
inherently subjective problem as we show in our work.  For the purposes of our
study, we focus exclusively on text-based toxic content, but attacks may also
extend to images and videos~\cite{zannettou2018origins}.

Previous studies have shown that some demographic cohorts in the United States
are more at-risk of receiving and reporting toxic content than
others~\cite{cowan2002effects,pew2017harassment}. For example, a survey by Pew
found that men were more likely to report experiencing offensive name calling
and physical threats, while women were more likely to experience sexual
harassment~\cite{pew2017harassment}. Beyond gender, Black adults were found to
report higher rates of name calling and purposeful
embarrassment~\cite{pew2017harassment}, while people who identify as LGBTQ+ were
three times as likely to report offensive name calling, physical threats, and
sexual harassment~\cite{datasociety2016harassment,blackwell2016lgbt}. Similarly
detailed demographic studies from various global perspectives are not yet
available. In order to ensure that automated detection works for all people,
including at-risk groups, we argue that it is critical to first understand how
different people perceive toxic content and how perceptions generalize
across Internet users.

\subsection{Detecting toxic content}
Security researchers and practitioners have proposed a multitude of
blocklist-based, machine learning, and natural language processing techniques to
detect toxic content. The simplest of these approaches rely on manually curated
lists of abusive words or users, such as HateBase's corpus of hate speech
related terms~\cite{hatebase}, or BlockTogether's list of
abusive Twitter accounts~\cite{jhaver2018online}. These provide targeted
protections against exact matches of terms or known abusers, but fail to generalize to
other types of toxic content, or in the context of blocklists, anonymous posts.

More sophisticated machine learning models include Yahoo's regression model
trained on a corpus of roughly 300,000 abusive comments with crowdsourced labels
that included hate speech, derogatory messages, and
profanity~\cite{nobata2016abusive}. Using a variety of NLP-based features, they
found their classifier could achieve an AUC of 0.90, though domain-specific
language and concept drift (\eg changes in abusive terms) degraded performance
over time. Since then, a variety of models have incorporated crowdsourced labels
such as Wikipedia moderation decisions~\cite{wulczyn2017ex, dixon2018measuring},
in-game conversations~\cite{blackburn2014stfu}, and social media
posts~\cite{chatzakou2017mean,davidson2017automated,dinakar2011modeling,van2015automatic,djuric2015hate,saravanaraj2016automatic}
to varying degrees of success. In another example, Founta et~al.\ leveraged
HateBase to build crowdsourced sublabels from participants for abusive tweets,
and then characterized a sample of Twitter data~\cite{founta2018large}.  Related
approaches have examined how to take a model trained for one community and apply
it to a separate community or site to avoid the cost of generating a labeled
training set~\cite{chandrasekharan2017bag}. Finally, several studies have
focused on latent annotator bias in
datasets~~\cite{razo2020investigating,wich2020investigating} and also
demonstrated that disagreements between raters for social tasks may explain why
classifiers excel on benchmarks but suffer in practice~\cite{gordon2021disagreement}.

Prominent models deployed at-scale today include Jigsaw's Perspective API, a
deep learning classifier for detecting toxic comments which is used by the New
York Times, Disqus, and other news sites for moderating toxic
comments~\cite{perspective-api}. Similarly, Instagram recently deployed a model
for nudging users away from posting comments that the classifier perceives as
harassment due to similar abusive text being reported in the
past~\cite{instagram-bullying}. We evaluate how these models
generalize across users in Section~\ref{section:comparisons}.


\vspace{-5pt}
\subsection{Other intervention strategies}
\looseness=-1
While our work focuses on how best to train classifiers to automatically detect
toxic content, researchers have also considered a variety of other strategies
for moderating toxic content. One example is building mechanisms into online
platforms to escalate conflicts to community tribunals who are empowered to
remove toxic content and take action against abusive users~\cite{riot-nature}.
Other examples include enabling bystanders to simply report toxic
content~\cite{difranzo2018upstanding}, or providing family and friends with
tools to assist in moderating toxic content on behalf of a
target~\cite{squadbox, blackwell2017classification}. All of these techniques
leverage community and context to overcome the limitations of automated
classification, but alone may fail to scale to the hundreds of millions of
interactions that happen online every day. Additionally, these systems cannot
relieve moderators of the emotional burden of reviewing toxic
content~\cite{trauma-floor}.

\subsection{Differentiation from prior work}
Prior work in evaluating automated toxicity classifiers has focused on either
investigating underlying bias in training data, such as flagging comments with
the word ``gay'' as hateful~\cite{davidson-racial,dixon2018measuring}, or shown
that classifiers are easily manipulated by substituting ``offensive'' words
while retaining semantic meaning~\cite{jain2018adversarial}. The focus of our
work is to first, understand how perspectives of toxic content change based on
individual experiences, and second, evaluate the impact these experiences have
on automated toxic content detection (Section~\ref{section:comparisons}). Prior
work identified certain groups to be at higher risk of online
harassment~\cite{pew2017harassment,datasociety2016harassment}, however, no work
has shown whether these experiences lead to differences in perception of toxic
content online. Closest to this is work by Cowan et~al.\, who investigated
perceptions of hate speech against three target groups on college campuses.
However, their study is limited in scale (N $<$ 500) and not specific to an
online context; our work focuses on a broader set of participants, focuses on
several categories of toxic content, and is more representative of online
discussion.  Furthermore, we investigate if implementing a personalized
filter---one that better captures the sentiment of participants by their
individual experiences---can improve toxicity detection.

\begin{table}[t]
    \footnotesize
\begin{tabular}{p{4cm}|rrr}
\toprule
			  		& \bf Offensive & \bf Hateful & \bf Toxic \\
                         		& N=72 &  N=74 & N=79 \\
\bf Theme raised by participants & $\kappa = 0.95$ & $\kappa = 0.98$ & $\kappa = 0.9$\\
\midrule
Insulting, demeaning, or derogatory 		& 42--44\%  & 55\%      & 58--62\%\\
Identity attack, hate speech, or racist 	& 33--35\%  & 39--41\%  & 33--34\%\\
Profane or obscene 				& 21\%      & 12\%      & 19\%\\
Threatening or intimidating 			& 11\%      & 11\%      & 16\%\\
Not constructive or off-topic 			& 3\%       & 0\%       & 9--11\%\\
\midrule
None of the above				& 29--32\%  & 20--22\%  & 19--20\%\\
\bottomrule
\end{tabular}

\caption{{\bf Interpretation of the Terms: Offensive, Hateful, and Toxic}---%
We find the term toxic resulted in the broadest interpretation for our rating task.}
\label{table:coding_themes}
\end{table}

\section{Methods}
\label{section:methodology}
\subsection{Survey instrument}
Our survey consisted of three parts: pre-exercise questions about
the participant's attitude towards technology and toxic content, an exercise where
the participant rated 20~comments from social media and community forums as
toxic or not, and finally, demographic and attention check questions.
We provide our full survey instrument in the Appendix. Our study was approved
by our institution's IRB.

\paragraph{Selecting terminology and comprehension.}
As a preliminary step, we first determined what terminology to use for our
rating task. An inherent challenge here is the ambiguity of the term \emph{toxic
content} or \emph{hate and harassment} and a lack of best practices from both
researchers and industry~\cite{pater2016characterizations}.

In the absence of common best practices, we ran a pilot study with $N=300$
participants recruited from Mechanical Turk to identify the terminology we
should use in our survey instrument. We asked each participant the open ended
question: ``When you see a post or comment, what do you look for to decide if
it's $<x>$?'', where $x$ was one of ``hateful'', ``offensive'', or ``toxic''.
We recruited $N=100$ participants per survey variant.  We did not use the term
``abusive'' as not to overload its meaning with other online abuse such as
for-profit cybercrime or unsafe content including drugs or self-harm. After
filtering for attention checks, we received a total of $N=225$ responses.

We reviewed each response and identified five emergent themes, detailed in
Table~\ref{table:coding_themes}. Two independent raters coded every response
according to these themes, with multiple themes possible per response.  Coding
achieved an interrater agreement Cohen's kappa $\kappa > 0.9$ for all three
variants, indicating strong agreement.\footnote{In the event that a rater
ascribed multiple themes to a single open ended response, we required both
raters to select the same set of themes to constitute agreement.} We found that
participants most often interpreted ``offensive'' to mean comments that were
insulting, profane, or an identity-based attack. Participants even more narrowly
construed the term ``hateful'' to mean comments that involved an
identity-related attack or insult.  On the other hand, ``toxic'' encompassed the
largest set of themes, where participants also considered whether a comment was
constructive or off-topic, and whether a comment was threatening. Based on our
findings, we adopted ``toxic'' as our final survey term to describe our rating
task to participants.

\paragraph{Determining the number of ratings per comment.}
Our survey instrument had to satisfy two competing goals: capturing a diverse
enough set of ratings to measure divergence among participants while also
maximizing the number of comments rated by participants to produce a
meaningfully-sized evaluation corpus. In order to identify how many ratings we
should solicit per comment, we ran a pilot survey where 100~participants rated a
fixed set of 200~manually curated comments where every comment was rated by
10~unique participants. Participants selected their rating on a five-point
Likert scale ranging from ``Not at all toxic'' to ``Extremely toxic''.

We then measured how quickly each comment's ratings converged to its average
toxicity score. In this context, we define the average toxicity score to be the
fraction of participants that labeled a comment as ``Moderately toxic'' or
greater per comment (the top three ratings of our Likert scale). The global
average is the average across all raters for each comment. We then measured the
number of participants required for the running average rating to fall within
10\% of the global average toxicity score per comment.

\begin{figure}[t]
    \centering
    \includegraphics[width=\columnwidth]{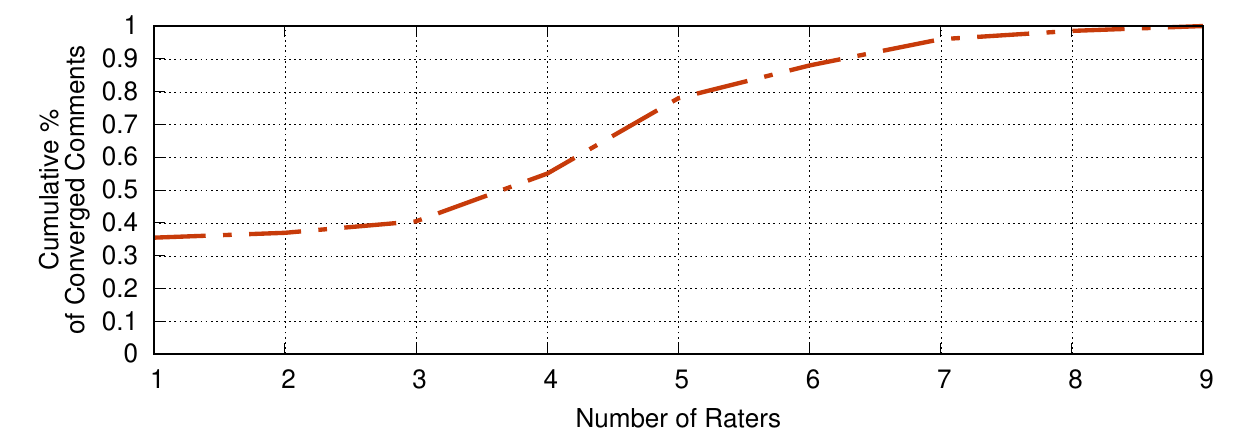}
    \caption{\textbf{Toxicity Convergence}---%
        We observe an inflection point where after five participants rate a
comment, the benefit of additional perspectives falls off.
    }
    \label{fig:toxicity_convergence}
\end{figure}

Figure~\ref{fig:toxicity_convergence} shows a CDF of the number of ratings
required for convergence for our pilot data. With only 2~ratings, 37\% of
comments had converged to their final distribution. However, with five ratings,
we found 78\% of the comments had converged to their final distribution with
each incremental participant adding only marginal improvements toward the global
average. As we needed to balance soliciting as many ratings as possible per
comment with the cost of doing so, we selected five participants to rate each
comment for this study.

\subsection{Sourcing potentially toxic content}
\begin{table}[t]
    \centering
    \footnotesize
    \begin{tabularx}{\columnwidth}{Xrrr}
        \toprule
        \bf Stride      &   \bf Aggregate Rating   &   \bf \% Agreement    & \bf {\% Final Dataset}\\
        \midrule
        0.0---0.1   &   Not toxic       &   90\%            &   5\% \\
        0.1---0.2   &   Not toxic       &   81.8\%          &   5\% \\
        0.2---0.3   &   Not toxic       &   80\%            &   5\% \\
        0.3---0.4   &   Not toxic       &   76.4\%          &   10\% \\
        0.4---0.5   &   Not toxic       &   71.4\%          &   10\% \\
        0.5---0.6   &   Not toxic       &   65.2\%          &   15\% \\
        0.6---0.7   &   Not toxic       &   68.3\%          &   15\% \\
        0.7---0.8   &   Toxic           &   65.2\%          &   20\% \\
        0.8---0.9   &   Toxic           &   76.4\%          &   10\% \\
        0.9---1.0   &   Toxic           &   80\%            &   5\% \\
        \bottomrule
    \end{tabularx}
    \caption{\textbf{Interrater Agreement per Stride}---%
        Although raters agree broadly for comments with either low or high
        toxicity scores, raters show minimal agreement when a comment is scored
        between 0.5---0.8. As such, we oversample these ranges for our dataset.
    }
    \label{table:interrater_agreement}
\end{table}

We sourced an initial corpus of 549,058 comments from Twitter, Reddit, and 4chan
for our study. We selected these platforms as they represent a diverse
cross-section of Internet users, are conversation driven, and contain varying
degrees of toxic behavior~\cite{twitter-report, reddit-hate, 4chan-hate}. All
data was collected between December 2019 and August 2020. Practically speaking,
each of these sites also provides a public API and real-time access to data.
While our dataset does not capture all types of conversations---such as private
discussions via messaging apps or ``walled gardens'' like Facebook---our
collection strategy avoids privacy constraints that would otherwise prevent
sharing content with random participants on crowdsourcing platforms.

Given the class imbalance inherent to each site, where benign content far
outweighs toxic content (with the exception of perhaps 4chan), a purely random
sampling approach would be prohibitively expensive to gather crowdsourced labels
for a sufficiently large volume of toxic content. Instead, we leveraged the
Perspective API \texttt{TOXICITY} model (discussed in detail in
Section~\ref{sec:background}) to build a stratified sample of potentially toxic
content.\footnote{Instagram does not provide a public API, thus we did not
consider it when building our dataset.} The API takes as input a sample of text
and returns a score between 0 and 1, describing the likelihood that an audience
would perceive the text to be toxic.

In order to identify which score ranges correlated with the largest rating
disagreement among participants, we ran a pilot
survey where 200~participants rated 800~comments, with 80~comments sourced from
each 0.1-stride between 0 and 1. For example, we selected 80~comments with a
toxicity score of 0---0.1, 80~comments with a score of 0.1---0.2, and so on.
Five independent participants rated each individual comment. We then measured
the interrater agreement for each stride as shown in
Table~\ref{table:interrater_agreement}. We found that participants broadly
agreed on comments that had a \texttt{TOXICITY} score of $<0.3$ or $>0.9$, with the least
agreement when a comment had a score of between of $0.5$ and $0.6$ and between
$0.7$ and $0.8$. A comment with a score of 0.5 might look like:

\begin{displayquote}
    ``I'm so sick of this mess. The Dems are not good because the
    Repubs are bad. The Repubs are not good when the Dems are bad. The enemy of your
    enemy can still be your enemy. \#BothPartiesSuck''
\end{displayquote}

Table~\ref{table:interrater_agreement} shows the final distribution of comments
we include per stride. Our dataset preferentially includes comments with lower
interrater agreement, however, we note that at least 5\% of comments are sampled
from each API stride. Our data distribution by source is 67\% Twitter comments,
15\% Reddit comments, and 18\% 4chan comments. We note our final dataset
contains at least 16,000 comments per platform. Our sampling skews towards
Twitter as we wanted to guarantee a fixed ratio of comments per stride while
maintaining a large corpus (N $>$ 100,000) but were limited by fraction of
comments available in each stride from 4chan and Reddit.

\subsection{Recruitment and validation}
We recruited participants for our final survey through Amazon Mechanical Turk to
``Participate in a survey about content online''. Previous studies have
validated the use of Mechanical Turk in security and privacy
contexts~\cite{redmiles2019well}. Given the scale of this work, we needed to
balance overall cost, fair compensation, and the goal of attracting a large and
diverse sample of workers across MTurk. After piloting, we decided to pay \$1
for completion. Participants took a median of 13 minutes to complete the task.
We only recruited participants with at least a 95\% approval
rating~\cite{redmiles2018asking} and restricted participants to residents of the
United States. All participants were over the age of 18. As our survey
instrument collects potentially sensitive demographic information (gender,
sexual orientation, race, and more), we provided an option to decline every
demographic question. As mentioned previously, our survey was approved by our
IRB.


In order to validate a participant's responses, we relied on an attention check
question at the end of the survey that asked participants to recall what term we
had used throughout the survey (\ie toxic). Additionally, we included an open
ended question asking participants to describe how they define toxic content
(akin to our pilot) and set a manually identified threshold on this response. We
solicited new participants until we reached our $n=5$ threshold per comment. Our
final dataset consists of \uniqueraters participants and \uniquecomments rated
comments.

\begin{table}[t]
    \centering
    \footnotesize
    \begin{tabularx}{\columnwidth}{Xrr}
        \toprule
        \bf Demographic &   \bf Cohort &   \bf \% Respondents \\
        \midrule
        Gender  &   Male        &   46\% \\
                &   Female      &   52\% \\
                &   Nonbinary   &   1\% \\
        \midrule
        Age     &   18 – 24     &   12\% \\
                &   25 – 34     &   40\% \\
                &   35 – 44     &   25\% \\
                &   45 – 54     &   13\%  \\
                &   55 – 64     &   7\% \\
                &   65+         &   3\% \\
        \midrule
        Race \& Ethnicity   &   Non-minority & 71\% \\
                            &   Minority & 29\% \\
        \midrule
        LGBTQ+ status       &   Not LGBTQ+ &   81\% \\
                            &   LGBTQ+     &   16\% \\
        \midrule
        Religion importance &   Not important   &   32\% \\
                            &   Not too important   &   12\% \\
                            &   Somewhat important  &   23\% \\
                            &   Very important  &   31\% \\
        \midrule
        Political attitude  &   Liberal         &   40\% \\
                            &   Independent     &   27\% \\
                            &   Conservative    &   27\% \\
        \midrule
        Parent		        & Yes & 52\% \\
                            & No & 47\% \\
        \bottomrule
    \end{tabularx}
    \caption{\textbf{Demographics of Respondents}---%
        Our recruitment strategy provided access to a diverse set of raters,
        including members of communities that are historically at-risk.  Not all
        percentages sum to 100\% due to some participants declining to provide
        demographic information.
    }
    \label{table:demographics}
\end{table}

Table~\ref{table:demographics} outlines the demographic distribution of our
participant pool. Participants were evenly split across men and women, with a
median age range of 25--34. Most participants identified as White, non-Hispanic
(71\%), and did not identify as a member of the LGBTQ+ community (81\%).
Attitudes towards religion were mixed with most participants either deeming
religion not important (32\%) or very important (31\%). Political attitudes were
mixed across Liberal, Independent, and Conservative participants. Our
participants also split evenly between parents and non-parents. Our sample does
not perfectly align to the US Census demographics for all demographic
cohorts~\cite{census}. However, our modeling results in Section~\ref{sec:rating}
control per demographic cohort and will stay consistent even if some cohorts are
over or under sampled. Overall, our recruitment provided access to a variety of
groups that historically are more likely to be the targets of toxic content. Our
dataset is available at
\url{https://data.esrg.stanford.edu/study/toxicity-perspectives}.

\subsection{Ethical considerations}
Given that our experiments expose participants to potentially toxic content, on
the Mechanical Turk description screen we included an initial warning that
described our rating task and the potential harms that might arise from
participating. We stated:

\begin{displayquote}
Risks related to this research include feeling targeted or potentially hurt by
viewing potentially toxic comments and recalling negative experiences in the
past regarding your personal experience with toxic comments online.
\end{displayquote}

{\noindent} At this point, participants could choose to accept the rating task
or simply move on without any exposure. After accepting the task, participants
consented to a longer agreement, that again reminded participants that they
would be exposed to toxic content multiple times. Additionally, our stratified
sampling approach avoided most egregious toxic content as detected by existing
automated classifiers, where there was unlikely to be any disagreement.  This is
in line with multiple prior studies that rely on crowdsourcing for toxic content
judgements~\cite{dixon2018measuring, blackburn2014stfu, jigsaw-kaggle}.
Overall, participants voluntarily saw a small number of potentially toxic
comments in a short session, most of which were rated to be only moderately
toxic and which are most in line with conversations that broadly occur on the
Internet.

\section{Toxicity, Filtering, and Removal Decisions}
\label{section:perspectives}
We examine how often participants deem a comment toxic and the frequency that
participants disagreed in rating the severity of toxicity per comment.
Additionally, we explore what classes of toxic content (\eg sexual harassment,
profanity) participants were most aligned in recognizing and ultimately their
personal beliefs of whether such content should be allowed online.

\subsection{Overall perceived comment toxicity}
\begin{figure}[t]
    \centering
    \includegraphics[width=\columnwidth]{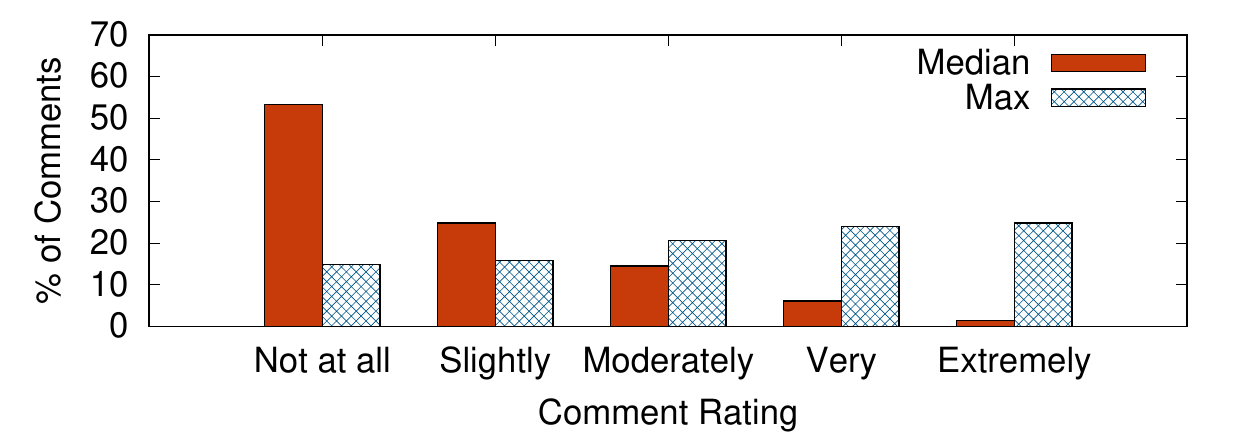}
    \caption{\textbf{Median and Maximum Toxicity Per Comment}--%
        The median toxicity rating per comment better captures the central
        tendency of raters compared to the max rating per comment.
    }
    \label{fig:median_max_bar}
\end{figure}

Each comment in our dataset includes five independent toxicity ratings drawn
from a Likert scale ranging from ``Not at all toxic'' to ``Extremely toxic.'' We
considered two strategies for aggregating these ratings into an overall
non-binary toxicity score per comment. The first, \emph{max rating}, selects the
maximum toxicity rating across all participants for a comment. The second,
\emph{median rating}, selects the median rating across all participants.
Figure~\ref{fig:median_max_bar} shows the distribution of toxicity scores via
these two metrics. Although the maximum rating achieves a relatively even
distribution across our Likert scale, this often results from a single outlier
among the five raters inflating the measured toxicity, compared to the median
rating. As such, we opted for the median rating and use it throughout this work
unless otherwise noted.

Overall, 53\% of comments in our dataset have a median rating of ``Not at all
toxic'', while only 1\% of comments have a median rating of ``Extremely toxic''.
An example of an ``Extremely toxic'' comment from 4chan was:

\begin{displayquote}
    ``They're like the polar opposite of limp-wristed smug douchebag homo's
    [sic] and liberals who buy a Prius and think they're better than everyone else''.
\end{displayquote}

{\noindent}Comments that were rated either ``Slightly toxic'' or ``Moderately
toxic'' make up 39.1\% of our labeled dataset. Examples from this category
include a Reddit comment like ``Kids with hoodies are going to be our future
criminals,'' and 4chan comments like ``Women can't be responsible for hiring
people. It is foolish to entrust hormonal women to the most important part of
the company.'' These examples show how even mildly toxic comments contain racism
and sexism---higher ratings merely represent the perceived intensity of the
harassment involved.

\subsection{Subcategories of toxic comments}
When participants rated a comment as ``Slightly toxic'' or higher, they also
provided an explanatory label for the toxic behaviors present in the comment.
Categories included ``Insult,'' ``Identity attack,'' ``Profanity,'' ``Threat,''
and ``Sexual harassment.'' Participants could select multiple such labels per
comment. As an example, the following comment was labeled as both a threat and
an identity attack: ``I think all you Deep State participants have done Enough
[sic] damage to this country.....go jump off the Brooklyn Bridge!''.  We derived
these labels from the themes surfaced by participants in our pilot study, adding
``sexual harassment'' as an additional theme and removing ``off topic'' due to
the lack of context provided to participants (see
Section~\ref{section:methodology}). We refer readers to the appendix for the
detailed instructions that we provided to participants on how to differentiate
these categories.

We present a breakdown of the perceived classes of toxic comments in our dataset
in Table~\ref{table:sublabel_by_rating}. Each column represents the fraction of
comments rated at each toxicity level that fell into each subcategory.  Overall,
insults are the most common type of toxic comment (67\%), followed by profanity
(52\%), and identity attacks (51\%). This is not necessarily an indication that
these are the most common toxic behaviors for sites in our sample, but rather
these are the toxic behaviors that raters identified. Participants also perceive
different sublabels as more or less toxic. For example, 85\% of ``Extremely
toxic'' comments involve an identity attack, whereas the same is true for only
57\% of comments rated ``Slightly toxic'' or lower. We also investigate the
reverse---which is the fraction of comments in each sublabel that fall into each toxicity
level, and find that participants perceive threats and sexual harassment as
``Extremely toxic'' (3.3\%, 3.7\% of comments respectively) at a higher rate
than identity attacks (2.9\%), profanity (2.6\%), and insults (2.3\%).

\begin{table}[t]
    \centering
    \small
    \begin{tabularx}{\columnwidth}{X|r|rrrr}
        \toprule
        \bf Category        	& \bf Overall
	& \rotatebox{90}{\parbox{1.6cm}{\bf Slightly Toxic}}
	& \rotatebox{90}{\parbox{1.6cm}{\bf Moderately Toxic}}
	& \rotatebox{90}{\parbox{1.6cm}{\bf Very Toxic}}
	& \rotatebox{90}{\parbox{1.6cm}{\bf Extremely Toxic}} \\
        \midrule
        Insult				&  67\%   &   76\%  &   85\%    & 89\%  & 89\% \\
        Profanity           &  52\%   &   59\%  &   69\%    & 74\%  & 78\% \\
        Identity attack	    &  51\%   &   57\%  &   70\%    & 79\%  & 85\% \\
        Threat	            &  31\%   &   30\%  &   44\%    & 54\%  & 59\% \\
        Sexual harassment   &  18\%   &   18\%  &   27\%    & 34\%  & 39\% \\
        \bottomrule
    \end{tabularx}
    \caption{\textbf{Categories of Toxic Content Recognized by Participants}--%
        Participants were most likely to perceive content as insulting or containing an
        identity attack, whereas sexual harassment and threats of violence or
        rape were less frequent.
    }
    \label{table:sublabel_by_rating}
\end{table}

\subsection{Frequency and intensity of disagreement}
While our overall score provides guidance on whether a plurality of participants
view a comment as toxic or not, in practice we are interested in how often
participants disagree and why. For example, of all comments with a median
toxicity of ``Not at all toxic'', only 28\% have uniform agreement among all
five raters. In order to measure diverging perspectives, we calculated the
variance of toxicity ratings for each comment. To do this, we treated each rating
as an ordinal value between 0 and 4. A variance of 0 indicates perfect agreement
for a comment. The maximum variance of 4.8 indicates two competing
groups (\eg two ``Extremely toxic,'' three ``Not at all toxic''). We opted for
variance over other multi-rater agreement metrics like Krippendorf's alpha or
Intra Class Correlation as we are interested in disagreement on individual
comments, not between raters.

Only 15\% of comments have a variance of 0, indicating all participants rated
the comment identically. In aggregate, the median variance of all rated comments
is 0.8. However, the spread of scores for comments rated as at least ``Slightly
toxic'' is larger, with a median variance of 1.3 per comment. As an example, the
comment from Twitter:

\begin{displayquote}
    ``At least REDACTED served, unlike you, a weirdo making memes online all day
    like a little lunatic.''
\end{displayquote}

{\noindent}had a variance of 1.3, with two raters finding the comment ``Very
toxic'', one rater finding the comment ``Moderately toxic'', one finding the
comment ``Slightly toxic'', and one rater not finding the comment toxic at all.
In contrast, 7.5\% of comments have a variance of 3.0 or greater, indicating
widespread disagreement. For example, the comment from Twitter:

\begin{displayquote}
     ``So you don't want money....
    Just free college, loan forgiveness, and (and I'm not sure how this is
    relevant) healthcare for veterans? I presume you believe only blacks were
    slaves? Also, your last sentence implies you believe all blacks were
    slaves...''
\end{displayquote}

{\noindent}had a variance of 3.2. Only 0.03\% of comments have a
variance of 4.8, which is the maximum amount.

Even when participants agree that a comment has some degree of toxicity, they
may still differ on why they feel a comment is toxic. Of comments that
participants uniformly deemed toxic, just 0.4\% had identical categories
assigned by all five participants. We quantify the degree of category
disagreement across our dataset using Fleiss' Kappa $\kappa$. This score
assesses how well a fixed number of raters place a subject into one of several
nominal categories---in our case, selecting the same set of categories (\eg
sexual harassment, insult) per comment. In order to arrive at an estimate, we
first calculated the $\kappa$ per block of comments\footnote{We note when
selecting five participants to rate a comment, the same five participants are
guaranteed to rate the same twenty comments in a random order, which enables us
to compare kappa values across participants per block of comments.} and then
calculated the global average.

The best group of five raters achieved a $\kappa = 0.57$, indicating only
moderate agreement~\cite{landis1977application}. The median group of raters
achieved a $\kappa = 0.10$, indicating low agreement. These findings illustrate
that participants are in general, more likely to agree on toxicity ratings than
on the justification for their decision.

\begin{figure}[t]
    \centering
    \includegraphics[width=\columnwidth]{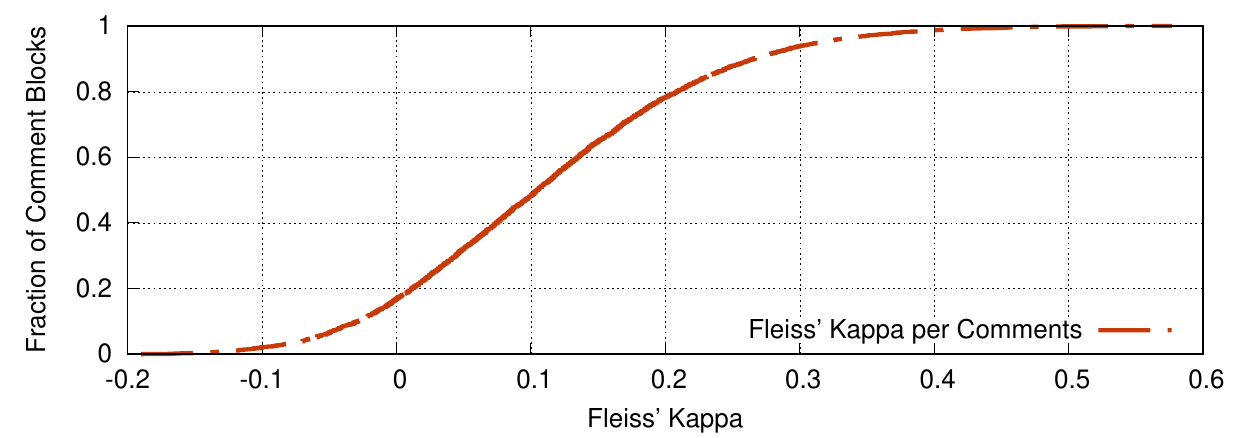}
    \caption{\textbf{Interrater Agreement for Subcategory Selection}--%
        Agreement between subcategories of raters is low, with a median
        interrater agreement score $\kappa$ of 0.10. The highest agreement
        between raters only reached 0.57 (moderate agreement), highlighting the
        difference between rater definitions of subcategories of toxic content.
    }
    \label{fig:kappa_cdf}
\end{figure}

\subsection{Filtering and removal recommendations}
Apart from the perceived toxicity of comments, we also asked participants to
make a decision for whether they personally would want to see each comment (\eg
personalized filtering), and whether the comment should be allowed online at all
(\eg global filtering). Of comments rated ``Slightly toxic'' or higher,
participants reported they would personally not want to see 37\% of comments.
We did not observe a strong distinction between personal filtering and global
filtering. In the event a participant felt personal filtering was appropriate,
they also felt that the comment should not be allowed online generally 70\% of
the time.  In the most extreme case, 30\% of participants would \emph{never}
remove a comment from an online platform---even for participants that rated at
least one comment as ``Extremely toxic'' (as 10\% of that 30\% of our
participants did).  That participants can recognize harassment but decline
intervention represents one of the fundamental conflicts between tackling toxic
content online and unfettered free speech.

We observe similar, competing perspectives when it comes to who participants
feel is the most responsible for addressing toxic content online. As part of our
pre-exercise questions, we asked participants whether they felt toxic content
was a problem and what party was most responsible for addressing toxic posts or
comments online. 42\% of participants felt toxic content was very frequently or
frequently a problem.  Another 51\% felt it was rarely or occasionally a
problem, while 5\% felt it was not an issue at all. Additionally, 47\% of
participants felt the onus of addressing toxic content was on the user who sent
the comment, compared to 27\% of participants who felt that the hosting platform
held the most responsibility.  This rift in beliefs---both for toxic content
being an issue online, and what party is responsible for solving it---represents
a challenge moving forward for tackling harassment online.

\section{Competing Perspectives of Toxicity}
\label{sec:rating}
\begin{table}[t]
    \centering
    \footnotesize
    \begin{tabularx}{\columnwidth}{X|ll|l}
	\toprule
	\bf Demographic & \bf Treatment & \bf Reference &	\bf Odds \\
	\midrule
	\multirow{2}{*}{Gender} 								& Female & Male &
    0.952 \\
	& Non-binary & Male & 0.707 \\
	\midrule
	\multirow{5}{*}{Age} & 18-24 & 35-44 & 1.238* \\
	                     & 25-34 & 35-44 & 1.227* \\
	                     & 45-54 & 35-44 & 0.972 \\
	                     & 55-64 & 35-44 & 0.980 \\
	                     & 65+   & 35-44 & 0.977 \\
        \midrule
        \multirow{1}{*}{Race \& Ethnicity}                                    &   Minority
&   Non-minority& 1.126* \\
	\midrule
	LGBTQ+	                &	LGBTQ+	&	Not LGBTQ+	&   1.644* \\
	\midrule
	\multirow{2}{*}{\shortstack[l]{Political\\affiliation}}	&	Conservative
    &	Liberal     &	1.024 \\
                                        		        &   Independent &
                                                        Liberal &   0.901* \\
	\midrule
	\multirow{3}{*}{\shortstack[l]{Importance\\of religion}}	&	Not too
    important	&	Not important	& 1.216* \\
								&	Somewhat important	&	Not important	&
                                1.572* \\
								&	Very important		&	Not important	&
                                1.840* \\
        \midrule
		Parent 			    & Is a parent & Not a parent & 1.330* \\
        \midrule
        \multirow{2}{*}{Education}  &   College &   High school &   1.139* \\
                                    &  Advanced degree  & High school & 1.365* \\
        \midrule
        \multirow{4}{*}{\shortstack[l]{Impact of\\technology\\on society}} & Very negative
        &   Neutral & 0.803* \\
                    & Somewhat negative &   Neutral &   0.870 \\
                    & Somewhat positive &   Neutral &   0.970 \\
                    & Very positive     &   Neutral &   1.142* \\
        \midrule
        \multirow{4}{*}{\shortstack[l]{Toxic content\\a problem?}} & Rarely    &
        Not a problem &   1.030 \\
            & Occasionally  &   Not a problem   &   0.958 \\
            & Frequently    &   Not a problem   &   1.029 \\
            & Very frequently   &   Not a problem   &   1.125* \\
        \midrule
        \multirow{4}{*}{\shortstack[l]{Party most\\responsible}}
	&  Law enforcement  & Bystander &   1.282* \\
        & Receiver          & Bystander  &   0.716* \\
        & Platform          & Bystander  &   0.706* \\
        & Sender            & Bystander  &   0.619* \\
	\midrule
	\multirow{2}{*}{\shortstack[l]{Witnessed\\toxic content}} & Yes & No &
    0.780* \\
	                                                          &     &    &      \\
	\midrule
	\multirow{2}{*}{\shortstack[l]{Target of\\toxic content}} & Yes & No &
    1.483* \\
								  &     &    & \\
		\bottomrule
    \end{tabularx}

        \caption{\textbf{Demographics, Experiences, and Opinions}---%
        We report the change in likelihood that a participant will flag a random
        comment as toxic, given a specific trait, in terms of odds. All values
        noted with an asterisk are significant with $p < 0.01$. See Appendix for
        model weights and exact significance values.\vspace{-10pt}
    }
    \label{table:model_results}
\end{table}

Given the frequency of disagreement among raters on what constitutes toxic
content, we explore potential explanatory variables stemming from a
participant's personal experiences, demographics, and opinions on whether toxic
content is a societal problem.

\subsection{Modeling participant decision making}
We treat each rating task per participant as a Bernoulli trial where a rating of
``Moderately toxic'' or higher indicates the participant found a comment toxic
(\eg a successful event, or 1), and all other ratings as benign (\eg failure, or
0). We then model the frequency of success across all labeling tasks as a
quasi-Binomial distribution $Y_i(n_i, \pi_i, \phi)$ using a logarithmic link
function. The model's parameters consist of categorical variables related to a
participant's age, gender, political affiliation, religious beliefs, LGBTQ+
affiliation, education, race and ethnicity, and parental status. The model also
incorporates whether a participant has previously witnessed toxic content online
or personally been the target of toxic content, whether the participant thinks
toxic content is an issue, and who is most responsible
for addressing toxic content.

Table~\ref{table:model_results} contains the results of our model. We report the
model's weights as the odds that a participant with a specific trait or
belief---after holding all other traits constant---will rate a comment randomly
drawn from our corpus as toxic. All results noted with an asterisk are
statistically significant with $p < 0.01$. While not shown in a table,
we repeat the same modeling process to also understand if any factors influence
a participant categorizing a toxic comment as any of our five subcategories of
toxic content. We report the full parameters of our models in the Appendix. We
discuss the results of our full analysis in detail below.

\subsection{Influence of personal experiences}
Overall, 77\% of participants reported having witnessed toxic content while
online. This aligns with a prior Pew study of personal experiences with online
harassment, which observed 73\% of Americans have observed online
harassment~\cite{pew2017harassment}. Conversely, 29\% of participants in our
study reported having been the target of toxic content.\footnote{Participants
answered both of these questions after the labeling task, which means their
answers may have been colored by the perceived toxicity, or lack thereof, of the
comments they labeled.} Both of these experiences exhibit a statistically
significant influence on toxicity ratings. Prior personal experience with being
the target of toxic content increases the odds of rating new content as toxic by
1.483~times.  These participants potentially empathize with others who might be
emotionally harmed by toxic content, and as such, take a stronger stance on what
behavior constitutes harassment. Conversely, prior experience with witnessing
toxic content decreases the odds of rating new content as toxic by 0.780~times.
These participants potentially view new toxic content through a comparative
lens, excusing abusive behavior that does not rise to the level of severity the
participant previously encountered. Our findings illustrate the importance of
understanding the experience of people who have been targets of harassment as
well as highlights the risk of desensitization.

\subsection{Influence of demographics}

\paragraph{Gender.}
We find no statistically significant differences between the odds that
non-binary, female, and male participants rate a comment as toxic. Furthermore,
female and male participants have nearly identical rates for identifying each
subcategory of toxic content. One exception is that the odds of a male
participant identifying a comment as threatening compared to female
participants increases by 1.158~times. One potential explanation is that men
report higher rates of physical threats and name calling compared to
women~\cite{pew2017harassment}, and may be more sensitive to those categories of
toxic content.

\paragraph{Age.}
We find that young participants in particular are more likely to flag comments
as toxic compared to older participants. Specifically, the odds of rating a
comment as toxic by people ages 18--34 increases 1.227--1.238 times compared to
participants aged 35--44. When comparing people 35--44 and groups of older
adults, we find no statistically significant difference between successive age
groups.  One possibility is that younger participants may be more represented on
the sites we sample from, and thus familiar with the slang or style of attacks
present. In line with previous studies~\cite{pew2017harassment,
datasociety2016harassment}, participants between the ages of 18--34 also
experienced online harassment at higher rates (27\%--30\% versus 20--24\%),
which may shape their opinion and sensitivity to toxic content.

\paragraph{LGBTQ+.}
A participant's LGBTQ+ identity plays a strong role in toxicity ratings.
Identifying as LGBTQ+ increases the odds of rating a comment as toxic by 1.644
times compared to participants who do not. Furthermore, LGBTQ+ participants were
far more likely to assign all subcategories to toxic comments---with threats
showing the largest increase in odds (1.865 times). LGBTQ+ participants are a
historically at-risk cohort for online
harassment~\cite{datasociety2016harassment} and so may be cognizant of toxic
behaviors, biases, and language that other participants fail to identify.

\paragraph{Importance of religion.}
Religion has one of the strongest influences on how participants perceive toxic
content. In particular, religion being ``Very important'' to a participant
increases the odds they rate a comment as toxic by 1.840 times. This impact still
holds even when a participant reports that religion is ``Not too important'',
where the odds of rating a comment as toxic increase by 1.216 times.  Similarly,
religious participants were far more likely assign all subcategories to toxic
comments---with profanity and threats showing the largest increase in odds
(1.604--1.878 times).

\paragraph{Parents.}
There is a small but statistically significant difference between the
perspectives of parents and non-parents. Being a parent
increases the odds of rating a toxic as comment by 1.330 times. Being a parent
also increased the odds of flagging sexually harassment (1.298 times) and profanity
(1.158 times). These differences are potentially influenced by content that
parents do not want their children to see online.

\paragraph{Race and Ethnicity.}
We find that belonging to a racial or ethnic minority plays only a small role in
influencing perspectives of toxic content, amounting to an increase in odds of
1.126 times compared to non-minority participants. Previous studies have shown
that minorities and non-minorities experience similar rates of online
harassment, but that when harassment occurs, people self-report it is more likely a result of
their race or ethnicity~\cite{pew2017harassment}.

\paragraph{Education and political affiliation.}
Compared to participants with only a high school education, the odds
participants with advanced degrees labeled comments as toxic increases 1.365
times, however, we find no similar relationship to those with college degrees
but no advanced degrees. Finally, we find that a participant's political
affiliation also has a small impact on the odds of identifying toxic content.
Notably, identifying as an independent decreases the odds of flagging toxic
content online by 0.901 times compared to liberal participants. These variations
may stem from the underlying content and discussions present in our dataset.

\subsection{Influence of technology beliefs}
Finally, we examine how attitudes towards technology and toxic content online
influence toxicity ratings. We find that, when participants feel that toxic
content is ``Very frequently'' a problem, the odds they flag content as toxic
increases by 1.125 times compared to others who feel toxic content is ``Not a
problem''.  Similarly, when participants feel that technology's role in peoples'
lives remains ``Very positive'', the odds they flag content as toxic increases
by 1.142 times compared to neutral participants. These participants potentially
have a lower threshold for what they deem to be toxic behavior, or feel a
greater obligation to address toxic content.

\section{Benchmarking Toxicity Classifiers}
\label{section:comparisons}
Given the influence of demographics, beliefs, and experiences on personal
toxicity ratings, we analyze how
well widely-deployed automated detection systems from Jigsaw and Instagram currently perform
in aggregate, per demographic cohort, and per individual.

\subsection{Perspective API}

\paragraph{Overall performance.}
As previously discussed, our dataset uses stratified sampling to oversample
potentially toxic comments with the highest rates of disagreement among
participants. We omit the vast majority of benign content on Twitter, Reddit,
and 4chan that would otherwise be present in a random sample. As such, it is
misleading to compare standard performance metrics (\eg accuracy,
precision-recall) across our entire dataset.  We control for this bias by
considering the accuracy of the Perspective API per \emph{stride} of our
sampling. As part of this, we convert every comment's rating distribution into a
binary verdict. We treat every comment with a median Likert score of
``Moderately toxic'' or higher as toxic and all other comments as benign. To
compute accuracy, we deem a perspective score of $>0.75$ as toxic and all other
comments as benign.

\begin{figure}[t]
    \centering
    \includegraphics[width=\columnwidth]{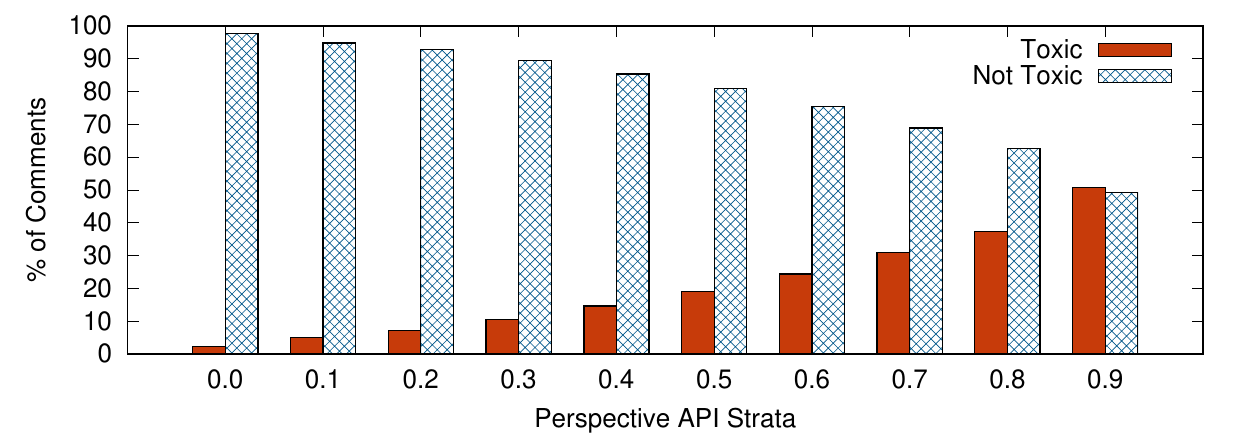}
    \caption{\textbf{Toxic/Benign Comment Distribution per Perspective API Stride}---%
        Higher Perspective API scores correlate with a larger fraction of
        toxic content, however, the fraction of toxic content per stride never
        exceeds the fraction of benign content.
    }
    \label{fig:accuracy_bar}
\end{figure}

Figure~\ref{fig:accuracy_bar} shows the fraction of toxic and benign content at
each stride of our dataset. The 0.1 stride includes all comments the Perspective
API gave a 0--10\% likelihood of being toxic, whereas the 0.9 stride includes
all comments with a 90-100\% likelihood of being toxic. While higher Perspective
API scores have monotonically increasing degrees of perceived toxicity, the
fraction of toxic content per stride is almost always smaller than fraction of
benign content, with the exception of the highest stride, where the labels are
roughly equal. Overall, we find only a weak correlation between our
participant's Likert ratings and the Perspective API ($r=0.39$, $p=0.0$). In
line with this, the accuracy for comments in the highest Perspective API stride
is only 51\%, indicating our participants disagreed with the Perspective rating
in 49\% of cases. As such, it appears that the Perspective API favors false
positives over false negatives. Such a balance is better suited for re-ranking
or informing moderation decisions as opposed to outright filtering.

\begin{table}[t]
    \centering
    \small
    \begin{tabularx}{\columnwidth}{Xrrr}
        \toprule
        \bf Stride      &   \bf \% Comments     &   \bf Accuracy    &   \bf RMSE \\
        \midrule
            0.0	&	5.0	    &   0.98    &   0.12\\
            0.1	&	5.0	    &	0.95    &   0.14\\
            0.2	&	5.0	    &   0.93    &   0.18\\
            0.3	&	10.0	&	0.90    &   0.24\\
            0.4	&	10.0	&	0.85    &   0.30\\
            0.5	&	15.0	&	0.81    &   0.36\\
            0.6	&	15.0	&	0.76    &   0.42\\
            0.7	&	20.0	&	0.50    &   0.48\\
            0.8	&	10.0	&	0.37    &   0.55\\
            0.9	&	5.0	    &	0.51    &   0.55\\
        \bottomrule
    \end{tabularx}
    \caption{\textbf{Accuracy and Root Mean Squared Error per Perspective API Stride}---%
        As Perspective API scores increase, we observe a larger divergence
        between what participants construe as toxic content versus the
        classifier. For comments that Perspective rates as 90\% likely to be
        toxic, our participants agreed only 45\% of the time.
    }
    \label{table:rmse}
\end{table}

\paragraph{Identifying divergent comments.}
To provide qualitative examples of divergence between participants and the
Perspective API, we isolated comments with the largest intensity of disagreement
between the Perspective API and our participants to understand the underlying
cause. This is best captured by a root mean square error (RMSE). In particular,
we first calculated the average rating across all participants for each comment
which we then normalized to a 0--1 scale. We relied on the original 0--1 score
of the Perspective API for comparison. We find that error is smallest for lower
scores of the Perspective API, indicating both the classifier and our
participants agree on what is not toxic. However, error increases as strides
increase, with error maximizing at the highest stride ($>=0.9$) as shown in
Table~\ref{table:rmse}. This result is consistent with our observations with
accuracy and indicates that the API struggles to match ground truth at high
decision thresholds.

We then selected the top 200 comments with the highest error between
participants and the Perspective API. Of these, we selected 100 where the
Perspective API skewed higher, and 100 where the Perspective API skewed lower.
Two independent coders reviewed each comment, came together to form a codebook
of high-level themes, and then independently rated every comment before
resolving any disagreement. Examples comments where Perspective and our
participants disagreed on toxicity ratings can be found in the
Appendix, Table~\ref{table:high_score} and Table~\ref{table:low_score}.

In some cases, the Perspective API is more sensitive to toxic content than our
participants. The majority of these comments included some form of profanity
that participants did not find toxic (82\%). For example, Perspective rates the
comment ``holy shit, you are perfect!'' at 0.83 while every rater found the
comment to be not at all toxic. Other sources of disagreement include differing
sensitivities to negativity (6\%), and scenarios where negative terms convey
emphasis (4\%). Examples include ``Song sucks'' and ``The stupid tongue! :D
Cutest thing'' respectively.  In effect, the classifier is incorrectly shaping
its decision based on negative or profane keywords alone.

In contrast, our participants also occasionally deemed content more toxic than
the Perspective API. Identity attacks against minorities, women, and the LGBTQ+
community made up 39\% of comments missed by the Perspective API. Another 17\%
of comments contained political messages that participants deemed toxic. For
example, participants rated the comment ``YOU'RE JUST LIKE YOUR BOSS AND THE
REST OF THE GOP A BUNCH OF!!!!! \#CLOWNS'' with a toxicity of 0.75, compared to
0.48 from the Perspective API\@. Other themes included adult content (11\%) and
threats of violence or rape (9\%). Additionally, despite Perspective API
regularly flagging profanity, minor grammatical changes such as the lack of
spaces in ``nofuckingbody'' resulted in score of 0.21, whereas adding spaces
results in a score of 0.93. Researchers have abused this sensitivity to minor
perturbations in text to construct adversarial examples that evade the
Perspective API~\cite{hosseini2017deceiving,grondahl2018all}.

\paragraph{Tuning classifiers to personal preferences.}
Our results indicate that a single definition of toxic content online
does not capture the varied experiences and opinions of Internet users. As such,
a one-size-fits-all model for abuse detection will likely not be able to capture the
toxicity preferences of all participants. Recent work from Google Jigsaw has
focused on allowing participants to ``Tune'' existing APIs to their own personal
preferences, simply by adjusting the Perspective API to a specific
threshold~\cite{tune-verge}. However, it is unclear how
effective this tuning strategy can be to end-users and where this mechanism
may fall short. We investigate the differences in accuracy and
precision for the optimal threshold for each individual participant compared to
the dataset in aggregate. Although our dataset is not a truly random sample of
Internet comments, comparing personal thresholds to the aggregate still provides
insight into the effectiveness of personal tuning.

To identify the optimal threshold for all ratings taken in aggregate, we first
convert each comment rating into a binary label. A comment rating has a positive
label if the participant personally did want to see the comment online, and a
negative label if they did not. We then sweep over all Perspective API decision
thresholds from 0--1 and identify the \emph{lowest} threshold that maximizes the
F1-score, which is the weighted average of the precision and recall. We find
that the optimal perspective API threshold for the aggregate dataset ranges from
0.18--0.49, all of which achieve a precision of 0.35 and an accuracy of 0.37.

\begin{figure}[t]
    \centering
    \includegraphics[width=\columnwidth]{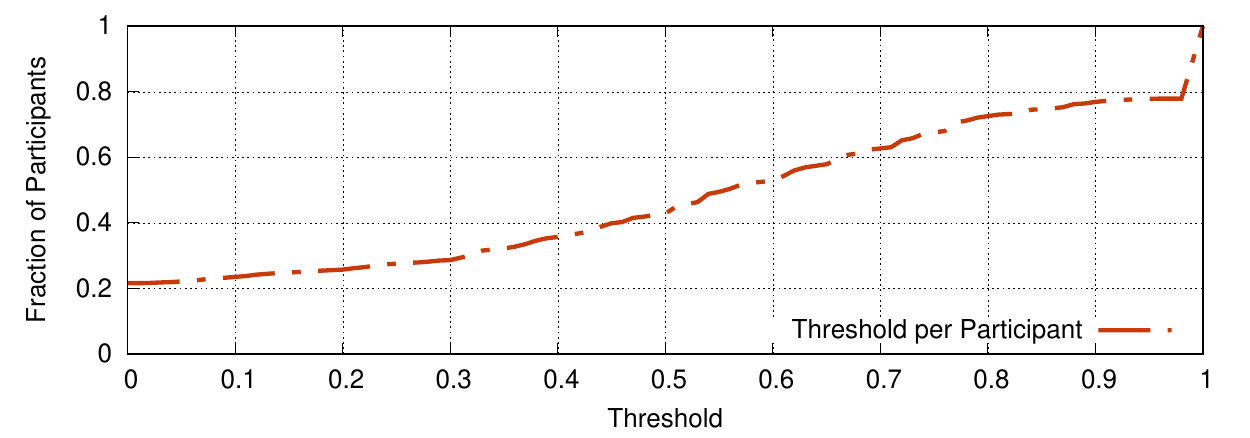}
    \caption{\textbf{Optimum Personalized Threshold per Participant}---%
        The threshold that maximizes classifier accuracy per participant is
        mixed. 21.6\% of participants are maximized at a threshold of 0.0,
        which amounts to labeling every comment as toxic. After tuning to
        personal thresholds, 71.5\% of participants achieved an accuracy greater
        than the overall classifier.
    }
    \label{fig:threshold_cdf}
\end{figure}

We perform the same analysis on an individual level, identifying a threshold
that maximizes the F1 score for each participant. If a participant did not
personally elect to remove any comments they encountered, we set their threshold
to the maximum possible value (1.0). Figure~\ref{fig:threshold_cdf} shows a
distribution of thresholds per individual. For 21.6\% of participants, their
maximal threshold is 0.0, suggesting that labeling every comment as toxic
maximizes both precision and recall. The median threshold is 0.61, resulting in
an average precision of 0.6 and an average accuracy of 0.68, an increase in
accuracy of 86\% compared to a one-size-fits-all classifier. Per this
personalized approach, 71.5\% of participants saw an improvement in accuracy over the
one-size-fits-all optimum accuracy. As such, more research is needed to
understand how best to quickly personalize models and how to gather ongoing
feedback in order to adjust model thresholds.

\paragraph{Tuning classifiers to demographic preferences.}
Given the differences between demographic groups outlined in
Section~\ref{sec:rating}, we also investigate the performance benefits for
tuning the Perspective model to broad demographic groups.
Table~\ref{table:demo_tuning} shows the maximum precision and accuracy when each
independent demographic group is tuned for separately. In contrast to
personalized tuning, where the performance increase was over twofold, we find
that demographic tuning in aggregate offers a smaller improvement over the
aggregate classifier---a 0--20.6\% increase in accuracy. Age-specific model
thresholds provided the best performance gain. These results highlight that even
within broad demographic groups, individual experiences and preferences take
more importance when making toxicity determinations online. Any cohort-based
model would need to account for multiple factors when designed and deployed.

\begin{table}[t]
    \centering
    \small
    \begin{tabularx}{\columnwidth}{X|rr|rr}
        \toprule
         &  \multicolumn{2}{c}{\bf Max Precision}    & \multicolumn{2}{c}{\bf
         Max Accuracy} \\
	\bf Demographic & \bf Value & \bf \% Change & \bf Value & \bf \% Change\\
        \midrule
        Religion        &   0.40 &  14.3\%          &   0.41 &   10.8\% \\
        Politics        &   0.37 &  5.7\%           &   0.37 &   0\% \\
        Age             &   0.44 &  25.7\%          &   0.44 &   20.6\% \\
        Gender          &   0.39 &  11.4\%          &   0.40 &   7.5\% \\
        Race            &   0.36 &  2.9\%           &   0.36 &   -2.7\% \\
        Parent          &   0.37 &  5.7\%           &   0.39 &   5.4\% \\
        LGBTQ+          &   0.36 &  2.9\%           &   0.37 &   0\% \\
        \bottomrule
    \end{tabularx}
    \caption{\textbf{Optimum Accuracy and Precision per Demographic}---%
        We show the maximum accuracy and precision when tuning the Perspective
        API per demographic cohort, as well as the percentage change from the
        one-size-fits-all model. We find that cohort-based models perform
        marginally better in some categories, but fall short of performance
        improvement from personalized models.
    }
    \label{table:demo_tuning}
\end{table}

\subsection{Instagram nudges}
In December 2019, Instagram rolled out a feature that nudges a user if they are
about to post a comment that is similar to comments that have been flagged for
abuse in the past. As a small experiment, we also compare how well the Instagram
classifier performs against our ground truth data. To do this, we first sampled
200 comments---150 of the most egregious ``toxic'' comments which have a median
toxicity rating of ``Very toxic'' or higher, and 50 ``benign'' comments that
have a median toxicity rating less than ``Slightly toxic''. We then manually
posted these comments to an Instagram account we controlled (with no audience),
noting which comments triggered the classifier.

Of the toxic comments, just 41 (27\%) triggered the Instagram classifier. In
manually inspecting these comments, we observed they were mostly identity-based
attacks (47\%), followed by a mix of adult content (15\%), profanity (7\%), and
threats (3\%). We followed a similar approach to labeling each comment with two
expert raters, but in the process, found there were no unifying themes that
might explain why toxic comments did not trigger detection. Aligned with this,
categories reported by our participants for our toxic sample included insults
(26\%), profanity (22\%), and identity attacks (21\%). The classifier never
triggered on a benign comment.  As such, a significant gap remains in the
classifier's ability to detect a wide variety of toxic comments.

\section{Discussion}
Based on our findings, we discuss potential best practices, pitfalls, and paths
forward for improving toxic content classifiers to better serve a diversity of
perspectives.

\paragraph{Best practices for crowdsourced labeling.}
\looseness=-1
During the development of our survey instrument, we were unable to identify any
best practices for developing crowdsourcing instruments that gather toxic
content ratings.  Previous studies used disparate terminology including
``abusive'', ``hateful'', ``offensive'', and ``toxic''.  Likewise, for
sublabeling tasks that involve categorizing toxic content into sexual
harassment, identity-based attacks, or insults, we were unable to find
terminology or a taxonomy that had previously been evaluated for rater
comprehension. Our experiments show that participants solicited from Mechanical
Turk in the United States best understood the meaning of ``toxic'' compared to
other terms, and that participants can identify at least five separate
categories of toxic content.  Given frequent rating disagreement between
participants, we also found that five ratings per comment resulted in the best
balance between minimizing crowdsourcing costs and achieving a high degree of
accuracy.  We believe this rating methodology can serve as a best practice when
crowdsourcing labels for toxic content, though we caution that any aggregate
label will not generalize to all demographic cohorts. Furthermore, our results
are limited to participants solicited from Mechanical Turk, and should be
validated with  participants from other crowdsourced platforms.

\paragraph{Towards personalized definitions of toxicity.}
Our results suggest that personalized tuning of one-size-fits-all models greatly
improves the accuracy per user compared to setting a global threshold for all
users. In particular, we found that per-user models increased the accuracy of
decisions by 86\%. These results suggest the feasibility of relying on a general
audience for training labels that users then tune to their personal preferences.
However, increasing classifier performance beyond this point will remain a
challenge without incorporating specific user feedback and examples. An
intermediate approach, where models generalize to specific single-trait
demographic cohorts rather than individuals, resulted in only a 0--20.6\%
improvement in accuracy, with age-specific models performing the best. In the
absence of personalization or user feedback, platforms might consider
increasingly sophisticated, community-based filters that take into account more
than just one demographic trait.


\paragraph{Measuring toxicity using existing classifiers.}
Recent studies in toxic content have begun to leverage toxicity classifiers as a
tool for measuring the prevalence of hate and harassment online, with additional
post-processing via rater
agreement~\cite{elsherief2018peer,hua2020towards,hua2020characterizing}.  Given
the variations in classifier accuracy across demographic cohorts and types of
sites, we caution against off-the-shelf usage of current classifiers without
such post-processing or additional calibration. Even at a Perspective toxicity
threshold of 0.9~or higher, our participants disagreed with the classifier's
verdict in 50\% of cases for the sites we measured.

\paragraph{Online Context.}
\looseness=-1
Our work does not incorporate the context that a comment is presented in. As
such, it may be challenging for a participant to pinpoint if a comment is toxic
versus simply sarcastic or joking. We selected this because toxicity detection
systems classify text without additional context, and we wanted to evaluate them
based on their current usage. Furthermore, users may have different responses to
toxic content when they see such context in-situ (e.g., the toxic content may be
targeted at an acquaintance). Some areas of future work include
understanding how perspectives change if participants are provided with
additional context when labeling, identifying if classifiers can be improved by
adding context during training, and measuring participant responses to
toxic content in-situ of browsing.

\section{Conclusion}
In this work, we built and deployed a survey instrument to 17,280~participants
across the United States and asked them about their perspectives on toxic
content online. We found that a participant's attitudes towards filtering toxic
content varies across a multitude of factors: their demographic background,
their personal experiences with harassment, and even their attitudes towards
technology and the state of toxic content online. Given these influences, we
showed how personalized tuning of independent thresholds for existing
classifiers can improve the accuracy of toxic detection performance by 86\% on
average, pointing to personalized models as a future area of research in toxic
content detection.  We have released our labeled toxicity dataset to
enable future work in this space and hope that our work presents paths forward
for improving toxic content classification for a diverse set of users.

\balance

{\footnotesize
\bibliographystyle{abbrv}
\bibliography{paper}}
\onecolumn
\setlength{\parindent}{0pt}
\footnotesize
\section*{Appendix---Survey instrument}

\paragraph{Initial consent form and university contact information}\\
I understand the consent form. I certify that I am 18 years old or older. By
clicking the “Yes” button to enter the survey, I indicate my willingness to
voluntarily take part in this study.\\
o Yes\\
o No\\
\hspace{2mm}

\paragraph{Pre-exercise questions}\\
Today we would like you to help us with this short survey and to review some
comments that have been posted online.\\

What types of sites do you use? [Checkbox] \\
o Social Networking (Facebook, Twitter) \\
o Video (YouTube, Twitch)\\
o News (CNN, Fox, NYT, WSJ)\\
o Community Forums (Reddit, Craigslist, 4chan)\\
o Email or messaging (Gmail, WhatsApp, Facebook Chat)\\
\hspace{2mm}

In general, how would you rate the impact of technology on people’s lives? [Radio]\\
o Very positive\\
o Somewhat positive\\
o Neutral – neither positive nor negative\\
o Somewhat negative\\
o Very negative\\
\hspace{2mm}

Based on your experience, toxic posts or comments are ... [Radio]\\
o Not a problem\\
o Rarely a problem\\
o Occasionally a problem\\
o Frequently a problem\\
o Very frequently a problem\\
\hspace{2mm}

Who is the most responsible for addressing toxic posts or comments? [Rank, 1-5]\\
o Law enforcement\\
o Users who observe the comment\\
o The user who receives the comment\\
o The website hosting the comment\\
o The user who sent the comment\\
\hspace{2mm}

What does the term “toxic comment” mean to you?\\
o Open Ended\\

\paragraph{Labeling exercise}\\
In this section, you will be shown twenty comments taken from various online
platforms. Some of these comments may contain toxic content. For each comment,
you will be asked a series of questions regarding your perspective on the
comment.\\

[Random comment displayed from Reddit, Twitter, or 4chan. This comment remains
visible throughout the subquestions below.]\\

How toxic do you personally find this comment? [Radio] \\
o Extremely toxic\\
o Very toxic\\
o Moderately toxic\\
o Slightly toxic\\
o Not at all toxic\\
\hspace{2mm}

[If any but not at all] Please select all the reasons why below: [Checkmark]\\
o It is threatening (ex. ``If you comment on my feed again, I will personally beat you up'')\\
o It is insulting (ex. ``You are an absolutely disgusting hypocrite'') \\
o It is profane (ex. ``Why the f*** not you heathen?'')\\
o It is an attack on a person or group based on their identity (ex. ``Women are stupid and belong in the kitchen'') \\
o It is sexual harassment (ex. ``All I want from you are nudes'') \\
o Other, please specify\\
\hspace{2mm}

Is this a type of comment you would want to see? [Radio]\\
o I would never want to see it\\
o It depends on the context\\
o This is fine for me to see\\
\hspace{2mm}

Should this comment be allowed online? [Radio]\\
o This comment should be removed\\
o It depends on the context\\
o This comment should be allowed\\
\hspace{2mm}

\paragraph{Post-exercise questions}\\
For the sites you use, have you ever seen comments similar to the ones we showed
you? [Radio]\\
o Yes\\
o No\\
\hspace{2mm}

Have you ever personally been the target of comments similar to the ones you
reviewed? [Radio]\\
o Yes\\
o No\\
\hspace{2mm}

Is there anything else you would like to tell us about toxic comments?\\
o [Open ended]\\
\hspace{2mm}

\paragraph{Demographic questions}\\
Which category below includes your age? [Radio]\\
o Under 18\\
o 18 - 24\\
o 25 - 34\\
o 35 - 44\\
o 45 - 54\\
o 55 - 64\\
o 65 or older\\
o Prefer not to say\\
\hspace{2mm}

Race [Checkbox]\\
o White\\
o Hispanic or Latino\\
o Black or African American\\
o Native American or American Indian\\
o Asian / Pacific Islander\\
o Other [open ended]\\
o Prefer not to say\\
\hspace{2mm}

What is your gender? [Radio] \\
o Female\\
o Male\\
o Nonbinary\\
o Prefer not to say\\
o Other [Open ended]\\
\hspace{2mm}

Would you describe yourself as transgender? [Radio] \\
o Yes\\
o No\\
o Prefer not to say\\
\hspace{2mm}

What is the highest degree or level of school that you have completed? [Radio] \\
o Less than high school degree\\
o High school graduate (high school diploma or equivalent including GED)\\
o Some college but no degree\\
o Associate degree in college (2-year)\\
o Bachelor's degree in college (4-year)\\
o Master's degree\\
o Doctoral degree\\
o Professional degree (JD, MD)\\
o Prefer not to say\\
o Other [Open ended]\\
\hspace{2mm}

Do you consider yourself to be: [Radio]\\
o Heterosexual or straight\\
o Homosexual\\
o Bisexual\\
o Prefer not to say\\
o Other [Open ended]\\
\hspace{2mm}

How important is religion in your life? [Radio]\\
o Not important\\
o Not too important\\
o Somewhat important\\
o Very important\\
o Prefer not to say\\
\hspace{2mm}

Generally speaking, do you consider yourself politically conservative, liberal,
independent, or something else? [Radio] \\
o Conservative\\
o Liberal\\
o Independent\\
o Other [Open ended]\\
o Prefer not to say\\
\hspace{2mm}

Are you a parent? [Radio]\\
o Yes\\
o No\\
o Prefer not to say\\
\hspace{2mm}

What term have we been using to describe comments in this survey? [Radio]\\
o Mean\\
o Abusive\\
o Bullying\\
o Toxic\\
\hspace{2mm}

\clearpage
\normalsize
\section*{Appendix---Divergent Comments}
\label{section:divergence}
We show examples of divergence between participant scores and the Perspective
API score in Table~\ref{table:high_score} and Table~\ref{table:low_score}.

\begin{table*}[h]
    \centering
    \small
    \begin{tabularx}{\textwidth}{Xp{4.5in}r}
        \toprule
        \bf Category    & \bf Sample & \bf \% Comments \\
        \midrule
        Profanity   & ``Damn cultural colloquialisms being all complicated and
stuff'' &  82\% \\
        Negative sentiment  &  ``Song sucks. But I vibe with how he mixed 3 of
Lil Wayne videos into one.'' &   6\% \\
        Inverted sentiment & ``The stupid tongue! :D Cutest thing'' & 4\% \\
        Adult & ``penis size shrinking in males exposed to contaminated
environments.''  & 2\%\\
        \midrule
        Other   &  &  6\% \\
        \bottomrule
    \end{tabularx}
    \caption{\textbf{High Perspective API Score, Low Groundtruth Score}--The
Perspective API over-emphasizes profanity, which participants rarely found
toxic.\vspace{10pt}}
    \label{table:high_score}
    \begin{tabularx}{\textwidth}{Xp{4.5in}r}
        \toprule
        \bf Category    & \bf Sample & \bf \% Comments \\
        \midrule
        Identity attack   &  ``We need to actively fight against racism by only
targeting white people and letting coloured people get away with crimes'' &  39\% \\
        Political   &   ``And you're getting closer to being locked up behind bars
\#pediphile \#RapistTrump'' & 17\% \\
        Adult & ``Hello World! Wanna Hookup; Trade Nude Pics? Visit ... & 11\% \\
        Threats & ``Store them in an unventilated room with hoses that run between
the room and your car's exhaust pipe. That'll solve your problem.'' & 9\% \\
        Profanity   &  ``the same person that invited you? nofuckingbody &  3\% \\
        Negative sentiment'' &  ``Sounds like you're a no one who's gonna die bitter
and alone and forgotten'' &  3\% \\
        \midrule
        Other  & &   18\% \\
        \bottomrule
    \end{tabularx}
    \caption{\textbf{Low Perspective API Score, High Groundtruth Score}--The
Perspective API currently misses nuanced identity-based attacks and political
attacks that participants deemed toxic.}
    \label{table:low_score}
\end{table*}

\newpage
\section*{Appendix---Regression Tables}
\label{section:regression_tables}
We present the full parameters and outputs for our quasi-binomial model that
predicts the likelihood of rating a random comment as toxic (1 or 0), and rating
a random comment they found as a particular sublabel (1 or 0). We also show
similar models for whether participants would rate a random toxic comment as an
insult, an identity attack, a threat, as profane, or as sexual harassment.  For
each model, we show the independent treatment group (either binary or
categorical depending on the variable), the reference group, the model
coefficient ($\beta$), error ($SE$), z-score ($z$), $p$-value, and the resultant
odds ratio ($OR$).

\begin{table}[ht]
\centering
\begin{tabular}{l|ll|rrrrr}
    \toprule
    \bf Demographic &   \bf Treatment   &   \bf Reference   &   $\beta$    &   $SE$ & $z$ & $Pr(>|z|)$ & $OR$ \\
    \midrule
    Gender  &   Female      &   Male   & -0.049 & 0.015 & -3.250 & 0.001 & 0.952 \\
    Gender  &   Nonbinary   &   Male    & -0.347 & 0.116 & -2.986 & 0.003 & 0.707 \\
    \midrule
    Age &   65 or older & 35 - 44  & -0.024 & 0.042 & -0.562 & 0.574 & 0.977 \\
    Age &   18 - 24 &   35 - 44 & 0.213 & 0.028 & 7.488 & 0.000 & 1.238 \\
    Age &   25 - 34 &   35 - 44 & 0.204 & 0.019 & 10.817 & 0.000 & 1.227 \\
    Age &   55 - 64 &   35 - 44 & -0.020 & 0.030 & -0.665 & 0.506 & 0.980 \\
    Age &   45 - 54 &   35 - 44 & -0.029 & 0.025 & -1.167 & 0.243 & 0.972 \\
    \midrule
    Race    &   Minority    &   Non-minority   & 0.119 & 0.016 & 7.277 & 0.000 & 1.126 \\
    \midrule
    LGBTQ+  &   LGBTQ+  &   Not LGBTQ+ & 0.497 & 0.020 & 25.225 & 0.000 & 1.644 \\
    \midrule
    Political affiliation   &   Independent     & Liberal   & -0.104 & 0.018 & -5.758 & 0.000 & 0.901 \\
    Political affiliation   &   Conservative    & Liberal  & 0.024 & 0.018 & 1.308 & 0.191 & 1.024 \\
    \midrule
    Religion    &   Not too important  &    Not Important     & 0.195 & 0.026 & 7.617 & 0.000 & 1.216 \\
    Religion    &   Somewhat important &    Not Important    & 0.453 & 0.021 & 21.947 & 0.000 & 1.572 \\
    Religion    &   Very important     &    Not Important    & 0.610 & 0.020 & 30.177 & 0.000 & 1.840 \\
    \midrule
    Parent  &   Yes     &   No  & 0.285 & 0.016 & 17.360 & 0.000 & 1.330 \\
    \midrule
    Education   &   College            &    High school     & 0.130 & 0.026 & 4.945 & 0.000 & 1.139 \\
    Education   &   Advanced degree    &    High school     & 0.311 & 0.030 & 10.325 & 0.000 & 1.365 \\
    \midrule
    Impact of Technology    &   Very negative     & Neutral      & -0.220 & 0.080 & -2.752 & 0.006 & 0.803 \\
    Impact of Technology    &   Somewhat negative & Neutral      & -0.140 & 0.032 & -4.357 & 0.000 & 0.870 \\
    Impact of Technology    &   Somewhat positive & Neutral      & -0.032 & 0.023 & -1.402 & 0.161 & 0.968 \\
    Impact of Technology    &   Very positive     & Neutral      & 0.133 & 0.025 & 5.318 & 0.000 & 1.142 \\
    \midrule
    Toxic Content a Problem?    &   Rarely a problem         & Not a problem  & 0.029 & 0.034 & 0.863 & 0.388 & 1.030 \\
    Toxic Content a Problem?    &   Occasionally a problem   & Not a problem  & -0.043 & 0.032 & -1.314 & 0.189 & 0.958 \\
    Toxic Content a Problem?    &   Frequently a problem     & Not a problem  & 0.028 & 0.033 & 0.848 & 0.397 & 1.029 \\
    Toxic Content a Problem?    &   Very frequently a problem& Not a problem  & 0.117 & 0.037 & 3.188 & 0.001 & 1.125 \\
    \midrule
    Party most responsible  &   Law Enforcement            & Bystander       & 0.248 & 0.035 & 7.093 & 0.000 & 1.282 \\
    Party most responsible  &   User who Receives          & Bystander       & -0.334 & 0.032 & -10.427 & 0.000 & 0.716 \\
    Party most responsible  &   Hosting Platform           & Bystander       & -0.348 & 0.028 & -12.481 & 0.000 & 0.706 \\
    Party most responsible  &   User who sent the comment  & Bystander       & -0.480 & 0.027 & -17.973 & 0.000 & 0.619 \\
    \midrule
    Witnessed Toxic Content &True   &   False   & -0.249 & 0.018 & -14.208 & 0.000 & 0.779 \\
    \midrule
    Experienced Toxic Content   &   True &  False & 0.394 & 0.017 & 23.547 & 0.000 & 1.482 \\
   \bottomrule
\end{tabular}
    \caption{\textbf{Toxicity Model}---%
        Logistic regression showing the likelihood a participant will flag a
        random comment as toxic.
    }
\end{table}

\begin{table}[ht]
\centering
\begin{tabular}{l|ll|rrrrr}
    \toprule
    \bf Demographic &   \bf Treatment   &   \bf Reference   &   $\beta$    &   $SE$ & $z$ & $Pr(>|z|)$ & $OR$ \\
    \midrule
    Gender  &   Nonbinary   &   Female    & -0.157 & 0.102 & -1.535 & 0.125 & 0.855 \\
    Gender  &   Male        &   Female    & 0.036 & 0.013 & 2.706 & 0.007 & 1.037 \\
    \midrule
    Age &   18 - 24 &   35 - 44     & 0.144 & 0.025 & 5.726 & 0.000 & 1.155 \\
    Age &   25 - 34 &   35 - 44     & 0.135 & 0.017 & 8.105 & 0.000 & 1.145 \\
    Age &   45 - 54 &   35 - 44      & -0.059 & 0.022 & -2.667 & 0.008 & 0.943 \\
    Age &   55 - 64 &   35 - 44     & -0.032 & 0.027 & -1.208 & 0.227 & 0.968 \\
    Age &   65 or older & 35 - 44   & -0.051 & 0.039 & -1.330 & 0.183 & 0.950 \\
    \midrule
    Race    &   Minority    &   Non-minority    & 0.072 & 0.014 & 5.031 & 0.000 & 1.075 \\
    \midrule
    LGBTQ+  &   LGBTQ+  &   Not LGBTQ+ & 0.327 & 0.017 & 19.152 & 0.000 & 1.387 \\
    \midrule
    Political affiliation   &   Independent     & Liberal   & -0.074 & 0.016 & -4.627 & 0.000 & 0.929 \\
    Political affiliation   &   Conservative    & Liberal   & 0.007 & 0.016 & 0.408 & 0.683 & 1.007 \\
    \midrule
    Religion    &   Not too important  &    Not Important     & 0.123 & 0.023 & 5.402 & 0.000 & 1.131 \\
    Religion    &   Somewhat important &    Not Important     & 0.275 & 0.018 & 14.911 & 0.000 & 1.316 \\
    Religion    &   Very important     &    Not Important     & 0.373 & 0.018 & 20.610 & 0.000 & 1.452 \\
    \midrule
    Parent  &   Yes     &   No  & 0.143 & 0.015 & 9.729 & 0.000 & 1.154 \\
    \midrule
    Education   &   College            &    High school     & 0.077 & 0.024 & 3.276 & 0.001 & 1.080 \\
    Education   &   Advanced degree    &    High school     & 0.152 & 0.027 & 5.651 & 0.000 & 1.164 \\
    \midrule
    Impact of Technology    &   Very negative     & Neutral      & -0.081 & 0.073 & -1.110 & 0.267 & 0.923 \\
    Impact of Technology    &   Somewhat negative & Neutral      & -0.034 & 0.029 & -1.167 & 0.243 & 0.967 \\
    Impact of Technology    &   Somewhat positive & Neutral      & 0.072 & 0.021 & 3.452 & 0.001 & 1.074 \\
    Impact of Technology    &   Very positive     & Neutral      & 0.121 & 0.023 & 5.340 & 0.000 & 1.128 \\
    \midrule
    Toxic Content a Problem?    &   Rarely a problem         & Not a problem  & 0.158 & 0.030 & 5.202 & 0.000 & 1.171 \\
    Toxic Content a Problem?    &   Occasionally a problem   & Not a problem  & 0.159 & 0.029 & 5.455 & 0.000 & 1.172 \\
    Toxic Content a Problem?    &   Frequently a problem     & Not a problem  & 0.205 & 0.030 & 6.900 & 0.000 & 1.228 \\
    Toxic Content a Problem?    &   Very frequently a problem& Not a problem  & 0.228 & 0.033 & 6.895 & 0.000 & 1.256 \\
    \midrule
    Party most responsible  &   Law Enforcement            & Bystander       & 0.104 & 0.030 & 3.481 & 0.001 & 1.110 \\
    Party most responsible  &   User who Receives          & Bystander       & -0.146 & 0.028 & -5.129 & 0.000 & 0.864 \\
    Party most responsible  &   Hosting Platform           & Bystander       & -0.161 & 0.025 & -6.488 & 0.000 & 0.851 \\
    Party most responsible  &   User who sent the comment  & Bystander       & -0.228 & 0.024 & -9.574 & 0.000 & 0.796 \\
    \midrule
    Witnessed Toxic Content &True   &   False   & -0.107 & 0.016 & -6.749 & 0.000 & 0.899 \\
    \midrule
    Experienced Toxic Content   &   True &  False & 0.225 & 0.015 & 15.111 & 0.000 & 1.252 \\
   \bottomrule
\end{tabular}
    \caption{\textbf{Insult Model}---%
            Logistic regression showing the likelihood a participant will flag a
            random toxic comment as an insult.
    }
\end{table}

\begin{table}[ht]
\centering
\begin{tabular}{l|ll|rrrrr}
    \toprule
    \bf Demographic &   \bf Treatment   &   \bf Reference   &   $\beta$    &   $SE$ & $z$ & $Pr(>|z|)$ & $OR$ \\
    \midrule
    Gender  &   Male        &   Female    & 0.016 & 0.015 & 1.086 & 0.278 & 1.016 \\
    Gender  &   Nonbinary   &   Female    & -0.066 & 0.114 & -0.578 & 0.563 & 0.936 \\
    \midrule
    Age &   18 - 24 &   35 - 44     & 0.081 & 0.029 & 2.819 & 0.005 & 1.084 \\
    Age &   25 - 34 &   35 - 44     & 0.074 & 0.019 & 3.920 & 0.000 & 1.077 \\
    Age &   45 - 54 &   35 - 44     & 0.024 & 0.025 & 0.988 & 0.323 & 1.025 \\
    Age &   55 - 64 &   35 - 44     & 0.059 & 0.030 & 1.983 & 0.047 & 1.061 \\
    Age &   65 or older & 35 - 44   & 0.165 & 0.041 & 3.995 & 0.000 & 1.180 \\
    \midrule
    Race    &   Minority    &   Non-minority    & 0.030 & 0.016 & 1.818 & 0.069 & 1.030 \\
    \midrule
    LGBTQ+  &   LGBTQ+  &   Not LGBTQ+ & 0.239 & 0.019 & 12.275 & 0.000 & 1.270 \\
    \midrule
    Political affiliation   &   Independent     & Liberal   & 0.005 & 0.018 & 0.254 & 0.800 & 1.005 \\
    Political affiliation   &   Conservative    & Liberal   & -0.006 & 0.019 & -0.345 & 0.730 & 0.994 \\
    \midrule
    Religion    &   Not too important  &    Not Important     & 0.151 & 0.025 & 5.959 & 0.000 & 1.163 \\
    Religion    &   Somewhat important &    Not Important     & 0.255 & 0.021 & 12.336 & 0.000 & 1.291 \\
    Religion    &   Very important     &    Not Important     & 0.267 & 0.021 & 12.997 & 0.000 & 1.305 \\
    \midrule
    Parent  &   Yes     &   No  & 0.101 & 0.017 & 6.097 & 0.000 & 1.106 \\
    \midrule
    Education   &   College            &    High school     & 0.099 & 0.027 & 3.677 & 0.000 & 1.104 \\
    Education   &   Advanced degree    &    High school     & 0.184 & 0.030 & 6.038 & 0.000 & 1.202 \\
    \midrule
    Impact of Technology    &   Very negative     & Neutral      & -0.146 & 0.082 & -1.795 & 0.073 & 0.864 \\
    Impact of Technology    &   Somewhat negative & Neutral      & -0.081 & 0.032 & -2.533 & 0.011 & 0.922 \\
    Impact of Technology    &   Somewhat positive & Neutral      & -0.028 & 0.023 & -1.218 & 0.223 & 0.973 \\
    Impact of Technology    &   Very positive     & Neutral      & -0.047 & 0.025 & -1.885 & 0.059 & 0.954 \\
    \midrule
    Toxic Content a Problem?    &   Rarely a problem         & Not a problem  & 0.152 & 0.036 & 4.274 & 0.000 & 1.165 \\
    Toxic Content a Problem?    &   Occasionally a problem   & Not a problem  & 0.225 & 0.034 & 6.602 & 0.000 & 1.252 \\
    Toxic Content a Problem?    &   Frequently a problem     & Not a problem  & 0.289 & 0.035 & 8.318 & 0.000 & 1.335 \\
    Toxic Content a Problem?    &   Very frequently a problem& Not a problem  & 0.304 & 0.038 & 7.933 & 0.000 & 1.355 \\
    \midrule
    Party most responsible  &   Law Enforcement            & Bystander       & 0.019 & 0.034 & 0.557 & 0.577 & 1.019 \\
    Party most responsible  &   User who Receives          & Bystander       & -0.154 & 0.032 & -4.769 & 0.000 & 0.857 \\
    Party most responsible  &   Hosting Platform           & Bystander       & -0.135 & 0.028 & -4.815 & 0.000 & 0.874 \\
    Party most responsible  &   User who sent the comment  & Bystander       & -0.201 & 0.027 & -7.462 & 0.000 & 0.818 \\
    \midrule
    Witnessed Toxic Content &True   &   False   & -0.057 & 0.018 & -3.165 & 0.002 & 0.945 \\
    \midrule
    Experienced Toxic Content   &   True &  False & 0.145 & 0.017 & 8.591 & 0.000 & 1.156 \\
   \bottomrule
\end{tabular}
    \caption{\textbf{Identity Model}---%
        Logistic regression showing the likelihood a participant
        will flag a random toxic comment as an identity attack.
    }
\end{table}

\begin{table}[ht]
\centering
\begin{tabular}{l|ll|rrrrr}
    \toprule
    \bf Demographic &   \bf Treatment   &   \bf Reference   &   $\beta$    &   $SE$ & $z$ & $Pr(>|z|)$ & $OR$ \\
    \midrule
    Gender  &   Male        &   Female    & 0.147 & 0.022 & 6.585 & 0.000 & 1.158 \\
    Gender  &   Nonbinary   &   Female    & -0.353 & 0.211 & -1.672 & 0.095 & 0.703 \\
    \midrule
    Age &   18 - 24 &   35 - 44     & 0.156 & 0.046 & 3.363 & 0.001 & 1.169 \\
    Age &   25 - 34 &   35 - 44     & 0.184 & 0.028 & 6.635 & 0.000 & 1.202 \\
    Age &   45 - 54 &   35 - 44     & -0.080 & 0.038 & -2.115 & 0.034 & 0.923 \\
    Age &   55 - 64 &   35 - 44     & -0.117 & 0.047 & -2.497 & 0.013 & 0.889 \\
    Age &   65 or older & 35 - 44   & -0.168 & 0.069 & -2.426 & 0.015 & 0.845 \\
    \midrule
    Race    &   Minority    &   Non-minority    & 0.073 & 0.024 & 3.083 & 0.002 & 1.075 \\
    \midrule
    LGBTQ+  &   LGBTQ+  &   Not LGBTQ+ & 0.624 & 0.026 & 24.275 & 0.000 & 1.865 \\
    \midrule
    Political affiliation   &   Independent     & Liberal   & -0.079 & 0.028 & -2.812 & 0.005 & 0.924 \\
    Political affiliation   &   Conservative    & Liberal   & 0.084 & 0.026 & 3.210 & 0.001 & 1.087 \\
    \midrule
    Religion    &   Not too important  &    Not Important     & 0.187 & 0.045 & 4.113 & 0.000 & 1.205 \\
    Religion    &   Somewhat important &    Not Important     & 0.477 & 0.035 & 13.801 & 0.000 & 1.610 \\
    Religion    &   Very important     &    Not Important     & 0.630 & 0.034 & 18.752 & 0.000 & 1.878 \\
    \midrule
    Parent  &   Yes     &   No   & 0.319 & 0.026 & 12.238 & 0.000 & 1.376 \\
    \midrule
    Education   &   College            &    High school     & 0.200 & 0.046 & 4.343 & 0.000 & 1.221 \\
    Education   &   Advanced degree    &    High school     & 0.508 & 0.049 & 10.272 & 0.000 & 1.662 \\
    \midrule
    Impact of Technology    &   Very negative     & Neutral      & -0.384 & 0.144 & -2.668 & 0.008 & 0.681 \\
    Impact of Technology    &   Somewhat negative & Neutral      & -0.291 & 0.057 & -5.154 & 0.000 & 0.747 \\
    Impact of Technology    &   Somewhat positive & Neutral      & -0.013 & 0.036 & -0.367 & 0.714 & 0.987 \\
    Impact of Technology    &   Very positive     & Neutral      & 0.273 & 0.038 & 7.262 & 0.000 & 1.313 \\
    \midrule
    Toxic Content a Problem?    &   Rarely a problem         & Not a problem  & -0.104 & 0.041 & -2.551 & 0.011 & 0.901 \\
    Toxic Content a Problem?    &   Occasionally a problem   & Not a problem  & -0.282 & 0.040 & -7.019 & 0.000 & 0.754 \\
    Toxic Content a Problem?    &   Frequently a problem     & Not a problem  & -0.214 & 0.042 & -5.123 & 0.000 & 0.807 \\
    Toxic Content a Problem?    &   Very frequently a problem& Not a problem  & -0.180 & 0.050 & -3.638 & 0.000 & 0.835 \\
    \midrule
    Party most responsible  &   Law Enforcement            & Bystander       & 0.015 & 0.040 & 0.365 & 0.715 & 1.015 \\
    Party most responsible  &   User who Receives          & Bystander       & -0.218 & 0.042 & -5.243 & 0.000 & 0.804 \\
    Party most responsible  &   Hosting Platform           & Bystander        & -0.360 & 0.037 & -9.642 & 0.000 & 0.698 \\
    Party most responsible  &   User who sent the comment  & Bystander       & -0.522 & 0.036 & -14.502 & 0.000 & 0.593 \\
    \midrule
    Witnessed Toxic Content &True   &   False   & -0.261 & 0.026 & -9.980 & 0.000 & 0.771 \\
    \midrule
    Experienced Toxic Content   &   True &  False & 0.465 & 0.025 & 18.391 & 0.000 & 1.592 \\
   \bottomrule
\end{tabular}
    \caption{\textbf{Threat Model}---%
        Logistic regression showing the likelihood a participant will flag a
        random toxic comment as a threat.
    }
\end{table}

\begin{table}[ht]
\centering
\begin{tabular}{l|ll|rrrrr}
    \toprule
    \bf Demographic &   \bf Treatment   &   \bf Reference   &   $\beta$    &   $SE$ & $z$ & $Pr(>|z|)$ & $OR$ \\
    \midrule
    Gender  &   Male        &   Female    & 0.109 & 0.015 & 7.246 & 0.000 & 1.115 \\
    Gender  &   Nonbinary   &   Female    & -0.310 & 0.132 & -2.354 & 0.019 & 0.734 \\
    \midrule
    Age &   18 - 24 &   35 - 44     & 0.175 & 0.029 & 6.039 & 0.000 & 1.191 \\
    Age &   25 - 34 &   35 - 44     & 0.135 & 0.019 & 7.120 & 0.000 & 1.144 \\
    Age &   45 - 54 &   35 - 44     & -0.006 & 0.025 & -0.255 & 0.799 & 0.994 \\
    Age &   55 - 64 &   35 - 44     & 0.038 & 0.030 & 1.272 & 0.203 & 1.039 \\
    Age &   65 or older & 35 - 44   & -0.095 & 0.044 & -2.149 & 0.032 & 0.909\\
    \midrule
    Race    &   Minority    &   Non-minority    & 0.027 & 0.016 & 1.676 & 0.094 & 1.028 \\
    \midrule
    LGBTQ+  &   LGBTQ+  &   Not LGBTQ+ & 0.280 & 0.019 & 14.552 & 0.000 & 1.323 \\
    \midrule
    Political affiliation   &   Independent     & Liberal   & -0.012 & 0.018 & -0.650 & 0.516 & 0.988 \\
    Political affiliation   &   Conservative    & Liberal   & 0.040 & 0.018 & 2.213 & 0.027 & 1.041 \\
    \midrule
    Religion    &   Not too important  &    Not Important     & 0.231 & 0.026 & 8.784 & 0.000 & 1.260 \\
    Religion    &   Somewhat important &    Not Important     & 0.424 & 0.021 & 20.029 & 0.000 & 1.528 \\
    Religion    &   Very important     &    Not Important     & 0.473 & 0.021 & 22.586 & 0.000 & 1.604 \\
    \midrule
    Parent  &   Yes     &   No  & 0.147 & 0.017 & 8.787 & 0.000 & 1.158 \\
    \midrule
    Education   &   College            &    High school     & 0.068 & 0.027 & 2.505 & 0.012 & 1.070 \\
    Education   &   Advanced degree    &    High school     & 0.171 & 0.031 & 5.597 & 0.000 & 1.187 \\
    \midrule
    Impact of Technology    &   Very negative     & Neutral      & -0.302 & 0.086 & -3.498 & 0.000 & 0.739 \\
    Impact of Technology    &   Somewhat negative & Neutral      & -0.121 & 0.033 & -3.708 & 0.000 & 0.886 \\
    Impact of Technology    &   Somewhat positive & Neutral      & -0.048 & 0.023 & -2.109 & 0.035 & 0.953 \\
    Impact of Technology    &   Very positive     & Neutral      & -0.073 & 0.025 & -2.901 & 0.004 & 0.930 \\
    \midrule
    Toxic Content a Problem?    &   Rarely a problem         & Not a problem  & 0.147 & 0.033 & 4.412 & 0.000 & 1.159 \\
    Toxic Content a Problem?    &   Occasionally a problem   & Not a problem  & 0.133 & 0.032 & 4.140 & 0.000 & 1.142 \\
    Toxic Content a Problem?    &   Frequently a problem     & Not a problem  & 0.103 & 0.033 & 3.135 & 0.002 & 1.109 \\
    Toxic Content a Problem?    &   Very frequently a problem& Not a problem  & 0.105 & 0.037 & 2.839 & 0.005 & 1.111 \\
    \midrule
    Party most responsible  &   Law Enforcement            & Bystander       & 0.062 & 0.032 & 1.925 & 0.054 & 1.064 \\
    Party most responsible  &   User who Receives          & Bystander       & -0.190 & 0.031 & -6.036 & 0.000 & 0.827 \\
    Party most responsible  &   Hosting Platform           & Bystander       & -0.176 & 0.027 & -6.423 & 0.000 & 0.839 \\
    Party most responsible  &   User who sent the comment  & Bystander        & -0.278 & 0.026 & -10.592 & 0.000 & 0.757 \\
    \midrule
    Witnessed Toxic Content &True   &   False   & -0.106 & 0.018 & -5.986 & 0.000 & 0.899 \\
    \midrule
    Experienced Toxic Content   &   True &  False & 0.182 & 0.017 & 10.766 & 0.000 & 1.200 \\
   \bottomrule
\end{tabular}
    \caption{\textbf{Profanity Model}---%
        Logistic regression showing the likelihood a participant will flag a
        random toxic comment as profane.
    }
\end{table}

\begin{table}[ht]
\centering
\begin{tabular}{l|ll|rrrrr}
    \toprule
    \bf Demographic &   \bf Treatment   &   \bf Reference   &   $\beta$    &   $SE$ & $z$ & $Pr(>|z|)$ & $OR$ \\
    \midrule
    Gender  &   Male        &   Female    & 0.107 & 0.025 & 4.275 & 0.000 & 1.112 \\
    Gender  &   Nonbinary   &   Female    & -0.316 & 0.220 & -1.437 & 0.151 & 0.729 \\
    \midrule
    Age &   18 - 24 &   35 - 44     & 0.149 & 0.050 & 2.987 & 0.003 & 1.160 \\
    Age &   25 - 34 &   35 - 44     & 0.142 & 0.031 & 4.539 & 0.000 & 1.152 \\
    Age &   45 - 54 &   35 - 44     & 0.019 & 0.041 & 0.475 & 0.635 & 1.020 \\
    Age &   55 - 64 &   35 - 44     & -0.098 & 0.052 & -1.868 & 0.062 & 0.907 \\
    Age &   65 or older & 35 - 44   & 0.062 & 0.071 & 0.873 & 0.383 & 1.064 \\
    \midrule
    Race    &   Minority    &   Non-minority    & -0.005 & 0.027 & -0.200 & 0.842 & 0.995 \\
    \midrule
    LGBTQ+  &   LGBTQ+  &   Not LGBTQ+  & 0.486 & 0.030 & 16.338 & 0.000 & 1.625 \\
    \midrule
    Political affiliation   &   Independent     & Liberal   & -0.033 & 0.030 & -1.091 & 0.275 & 0.967 \\
    Political affiliation   &   Conservative    & Liberal   & -0.055 & 0.030 & -1.846 & 0.065 & 0.946 \\
    \midrule
    Religion    &   Not too important  &    Not Important     & 0.208 & 0.047 & 4.455 & 0.000 & 1.232 \\
    Religion    &   Somewhat important &    Not Important     & 0.443 & 0.037 & 12.122 & 0.000 & 1.557 \\
    Religion    &   Very important     &    Not Important     & 0.515 & 0.036 & 14.293 & 0.000 & 1.674 \\
    \midrule
    Parent  &   Yes     &   No   & 0.261 & 0.029 & 9.130 & 0.000 & 1.298 \\
    \midrule
    Education   &   College            &    High school     & 0.197 & 0.049 & 4.023 & 0.000 & 1.217  \\
    Education   &   Advanced degree    &    High school     & 0.278 & 0.054 & 5.151 & 0.000 & 1.320 \\
    \midrule
    Impact of Technology    &   Very negative     & Neutral      & -0.250 & 0.149 & -1.678 & 0.093 & 0.779 \\
    Impact of Technology    &   Somewhat negative & Neutral      & -0.094 & 0.056 & -1.664 & 0.096 & 0.910 \\
    Impact of Technology    &   Somewhat positive & Neutral      & 0.005 & 0.039 & 0.119 & 0.905 & 1.005 \\
    Impact of Technology    &   Very positive     & Neutral      & 0.039 & 0.042 & 0.928 & 0.354 & 1.040 \\
    \midrule
    Toxic Content a Problem?    &   Rarely a problem         & Not a problem  & 0.043 & 0.052 & 0.818 & 0.413 & 1.044 \\
    Toxic Content a Problem?    &   Occasionally a problem   & Not a problem  & 0.064 & 0.050 & 1.277 & 0.202 & 1.067 \\
    Toxic Content a Problem?    &   Frequently a problem     & Not a problem  & 0.111 & 0.052 & 2.141 & 0.032 & 1.117 \\
    Toxic Content a Problem?    &   Very frequently a problem& Not a problem  & 0.086 & 0.059 & 1.454 & 0.146 & 1.090 \\
    \midrule
    Party most responsible  &   Law Enforcement            & Bystander       & -0.036 & 0.049 & -0.735 & 0.462 & 0.965 \\
    Party most responsible  &   User who Receives          & Bystander       & -0.162 & 0.049 & -3.305 & 0.001 & 0.850 \\
    Party most responsible  &   Hosting Platform           & Bystander        & -0.249 & 0.043 & -5.726 & 0.000 & 0.780 \\
    Party most responsible  &   User who sent the comment  & Bystander       & -0.315 & 0.042 & -7.574 & 0.000 & 0.730\\
    \midrule
    Witnessed Toxic Content &True   &   False   & -0.138 & 0.029 & -4.716 & 0.000 & 0.871 \\
    \midrule
    Experienced Toxic Content   &   True &  False & 0.237 & 0.028 & 8.413 & 0.000 & 1.267 \\
   \bottomrule
\end{tabular}
    \caption{\textbf{Sexual Harassment}---%
        Logistic regression showing the likelihood a participant will flag a
        random toxic comment as sexual harassment.
    }
\end{table}

\end{document}